\documentclass{article}
\usepackage{graphicx}
\usepackage{xcolor}
\usepackage{amsmath, amsthm, amssymb}
\usepackage{times}

\newcommand{\bmp}{\begin{minipage}}
\newcommand{\emp}{\end{minipage}}
\newcommand{\Rt}{\mbox{\it R}^{\top}}
\newcommand{\Rb}{\mbox{\it R}^{\bot}}
\newcommand{\Ra}{\mbox{\it Set}_R}
\newcommand{\SL}{\mbox{\it SL}}
\newcommand{\SR}{\mbox{\it SR}}
\newcommand{\PopL}{\mbox{\it Pop}_L}
\newcommand{\PopR}{\mbox{\it Pop}_R}
\newcommand{\PushLR}{\mbox{\it Push}_{LR}}
\newcommand{\Id}{\mbox{\it Id}}
\newcommand{\tp}{\mbox{\it top}}
\newcommand{\LR}{\mbox{\it LR}}
\newcommand{\M}{\mbox{\it MinMax}}

\newcommand{\LpRp}{L^+R^+}
\newcommand{\LpRm}{L^+R^-}
\newcommand{\LmRm}{L^-R^-}
\newcommand{\LmRp}{L^-R^+}
\newcommand{\FindL}{\mbox{\it Find}_L}
\newcommand{\mins}{\mbox{\it m}_s}
\newcommand{\maxs}{\mbox{\it M}_s}

\newcommand{\Pm}{P^{-1}}
\newcommand{\BL}{\mbox{\it l}}
\newcommand{\BR}{\mbox{\it r}}

\newcommand{\bs}{\mbox{\it l}_s}
\newcommand{\Bs}{\mbox{\it r}_s}

\newcommand{\out}{{Filter}}

\newcommand{\PK}{\mathcal{P}}

\newcommand{\next}{\mbox{\it next}}

\newcommand{\W}{W}
\newcommand{\Precc}{\mbox{\it Prec}}
\newcommand{\Succ}{\mbox{\it Succ}}

\newcommand{\As}{A^{\it sep}}

\newcommand{\Od}{\mathcal{O}}
\newcommand{\B}{\mathcal{B}}
\newcommand{\bb}{b}
\newtheorem{thm}{Theorem}
\newtheorem{cor}{Corollary}
\newtheorem{fait}{Claim}
\newtheorem{rmk}{Remark}
\newtheorem{defin}{Definition}
\newtheorem{ex}{Example}

\newcommand{\br}{\begin{rmk}\rm}
\newcommand{\er}{\end{rmk}}
\newcommand{\bdefin}{\begin{defin}\rm}
\newcommand{\edefin}{\end{defin} }
\newcommand{\bex}{\begin{ex}\rm}
\newcommand{\eex}{\end{ex}}

\newcommand{\bthm}{\begin{thm}}
\newcommand{\ethm}{\end{thm}}
\newcommand{\bcor}{\begin{cor}}
\newcommand{\ecor}{\end{cor}}
\newcommand{\bfn}{\begin{fait}}
\newcommand{\efn}{\end{fait}}

\usepackage{algorithm}
\usepackage{algorithmic}



\renewcommand{\Box}{\rule{1.5mm}{3mm}}

\begin{document}


\begin{center}
{\bf\large Extending Common Intervals Searching from Permutations to Sequences }\\

%

\vspace*{1cm}

Irena Rusu

L.I.N.A., UMR 6241, Universit\'e de Nantes, 2 rue de
la Houssini\' ere,\\

 BP 92208, 44322 Nantes, France
\end{center}


\vspace*{1cm}

\hrule
\vspace{0.3cm}

\noindent{\bf Abstract}

Common intervals have been defined as a modelisation of gene clusters in genomes represented either as permutations
or as sequences. Whereas optimal algorithms for finding common intervals in permutations exist even for an arbitrary 
number of permutations,
in sequences no optimal algorithm has been proposed yet even for only two sequences. Surprisingly enough, when sequences are reduced to
permutations, the existing algorithms perform far from the optimum, showing that their performances are not
dependent, as they should be, on the structural complexity of the input sequences.

In this paper, we propose to characterize the structure of a sequence by the number $q$ of different dominating orders
composing it (called the {\em domination number}), and to use a recent algorithm for permutations in order to devise a new algorithm for two sequences.
Its running time is in $O(q_1q_2p+q_1n_1+q_2n_2+N)$, where $n_1, n_2$ are the sizes of the two sequences, $q_1,q_2$ are their
respective domination numbers, $p$ is the alphabet size and $N$ is the number of solutions to output.
This algorithm performs better as $q_1$ and/or $q_2$ reduce, and when the two sequences are reduced to permutations (i.e.
when $q_1=q_2=1$) it has the same running time as the best algorithms for permutations. It is also the first algorithm for sequences whose running time involves the parameter size of the solution. As a counterpart, when $q_1$ and $q_2$ are of $O(n_1)$ and $O(n_2)$ respectively, the algorithm is less efficient than other approaches.

\bigskip
\hrule 

\section{Introduction}

One of the main assumptions in comparative genomics is that a set of genes
occurring in neighboring locations within several genomes represent functionally related genes 
\cite{galperin2000s,lathe2000gene,tamames2001evolution}. Such clusters of genes 
are then characterized by a highly conserved gene content, but a possibly different order
of genes within different genomes. Common intervals have been defined  to
model clusters \cite{UnoYagura}, and have been used since to detect 
clusters of functionally  related genes \cite{overbeek1999use, tamames1997conserved}, 
to compute similarity  measures between genomes \cite{BergeronSim, AngibaudHow} 
and to predict protein functions \cite{huynen2000predicting, von2003string}. 

Depending on the representation of genomes in such applications, allowing or not the
presence of duplicated genes, comparative genomics requires for finding common intervals
either in sequences or in permutations over a given alphabet. Whereas the most
general - and thus useful in practice - case is the one involving sequences, the easiest
to solve is the one involving permutations. This is why, in some approaches \cite{AngibaudApprox, 
angibaud2006pseudo}, sequences are reduced to permutations by renumbering the copies of the same gene
according to evolutionary based hypothesis. Another way to exploit the performances of
algorithms for permutations in dealing with sequences is to see each sequence as a combination
of several permutations, and to deal with these permutations rather than with the sequences.
This is the approach we use here.

In permutations on $p$ elements, finding common intervals may be done in $O(Kp+N)$ time where $K$ is the number
of permutations and $N$ the number of solutions, using several algorithms proposed in the literature \cite{UnoYagura, BergeronK, heber2011common, IR2013}.
In sequences (see Table \ref{table:algos}), even when only two sequences $T$ and $S$ of respective sizes $n_1$ and $n_2$ are considered, the best solutions take quadratic time. 
In a chronological order, the first algorithm is due to Didier \cite{didier2003common} and performs in 
$O(n_1n_2\log n_2)$ time and $O(n_1+n_2)$ space. Shortly later, Schmidt and Stoye \cite{schmidt2004quadratic} propose 
an $O(n_1n_2)$ algorithm which needs $O(n_1n_2)$ space, and note that Didier's algorithm may benefit
from an existing result to achieve $O(n_1n_2)$ running time whereas keeping the linear space. Both
these algorithms use $T$ to define, starting with a given element of it, growing intervals of $T$ 
with fixed leftpoint and variable rightpoint, that are searched for into $S$. Alternative
approaches attempt to avoid multiple searches of the same interval of $T$,
due to multiple locations, by efficiently computing all intervals in $T$ and all intervals in $S$ 
before comparing them. The best running time reached by such an algorithm is in $O(n_1+n_2+l_1\log  p+l_2\log  p)$,
obtained by merging the fingerprint trees proposed in \cite{kolpakov2008new},
where $l_1$ (respectively $l_2$) is the number of maximal locations of the intervals in $T$
(respectively $S$), and $p$ is the size of the alphabet. The value $l_1$ (and similarly for $l_2$)
is in $\Omega(p^2)$ and does not exceed $n_1p$.

The running times of all the
existing algorithms have at least two main drawbacks: first, they do not involve at all the number $N$ of 
output solutions; second, they insufficiently exploit the particularities of 
the two sequences and, in the particular case where the sequences are
reduced to permutations, need quadratic time instead of the optimal $O(p+N)$ time for two permutations on $p$ elements.
That means that their performances insufficiently depend both on the inherent complexity of the input
sequences, and on the amount of results to output.
Unlike the algorithms dealing with permutations, the algorithms for sequences lack of criteria allowing
them to decide when the progressive generation of a candidate must be stopped, since 
it is useless.  This is the reason 
why their running time is independent of the number of output solutions. This is also the reason
why when sequences are reduced to permutations the running time is very unsatisfactory.

\begin{table}
\begin{center}\scalebox{0.79}{
\begin{tabular}{|l||c|c|c|c|c|}
 \hline
Sequence type& Didier \cite{didier2003common}&Schmidt and Stoye\cite{schmidt2004quadratic}&Kolpakov and Raffinot \cite{kolpakov2008new}& Our algorithm\\ \hline\hline
Seq. vs. Seq.&$O(n_1n_2)$&$O(n_1n_2)$&$O(n_1+n_2+l_1\log  p+l_2\log  p)$ &$O(q_1q_2p+q_1n_1+q_2n_2+N)$\\
Perm. vs. Seq.&$O(pn_2)$&$O(pn_2)$&$O(n_2+p^2\log  p+l_2\log  p)$&$O(q_2n_2+N)$\\
Perm. vs. Perm.&$O(p^2)$&$O(p^2)$&$O(p^2\log  p)$&$O(p+N)$\\ \hline\hline
Memory space &$O(n_1+n_2)$&$O(n_1n_2)$&$O(n_1+n_2+l_1\log  p+l_2\log  p)$&$O(n_1+n_2)$\\
\hline
\end{tabular}}
\end{center}
\caption{\small Running time for the existing algorithms when (1) the input sequences are as general as possible
(lengths $n_1$ and $n_2$), when (2) one
of them is a permutation (lengths $n_1=p=|\Sigma|$ and $n_2\geq p$), and when (3) both are permutations (lengths
$n_1=n_2=p=|\Sigma|$).  The running time of Didier's algorithm is updated according to
Schmidt and Stoye's remark. On the last line, we add the memory space needed by each algorithm ($n_1=p$
and $n_1=n_2=p$ respectively hold for the second and third case). Parameters $l_1,l_2, q_1, q_2$ are
described in the text.}
\label{table:algos}
\end{table}

The most recent optimal algorithm for permutations \cite{IR2013} proposes a general framework for efficiently 
searching for common intervals and all of their known subclasses in $K$ permutations, and has a twofold advantage,
not proposed by other algorithms. First, it permits an easy and efficient selection of the common intervals to 
output based on
two types of parameters. Second, assuming one permutation has been renumbered to be the identity permutation,
it outputs all common intervals with the same minimum value together 
and in increasing order of their maximum value. We use here these properties to propose a new algorithm for finding common intervals in two 
sequences. Our algorithm strongly takes into account the structure of the input sequences, expressed by the number
$q$ of different dominating orders (which are permutations) composing the sequence ($q=1$ for permutations). Consequently, it has a complexity depending both on this structure and 
on the number of output solutions. It runs in optimal $O(p+N)$ time for two permutations on $p$ elements, 
is better than the other algorithms 
for sequences composed of few dominating orders and, as a counterpart, it performs less well as the number of composing dominating orders grows. 

The structure of the paper is as follows. In Section \ref{sect:preliminaries} we define the main notions,
including that of a dominating order, and give the results allowing us a first simplification of the problem. 
In Section \ref{sect:approach} we propose our approach for finding common intervals in two sequences based
on this simplification, for which we describe the general lines. In Sections \ref{sect:find}, \ref{sect:retrieve} and \ref{sect:common} we develop each of these general lines and prove correctness and complexity results.
Section \ref{sect:conclusion} is the conclusion.

\section{Preliminaries} \label{sect:preliminaries}

Let $T$ be a sequence of length $n$ over an alphabet $\Sigma:=\{1, 2, \ldots, p\}$. We denote the length of $T$ by $||T||$,
the set of elements in $T$ by $Set(T)$, the element of $T$ at position $i$, $1\leq i\leq n$, by $t_i$ and 
the subsequence of $T$ delimited by positions $i,j$ (included), with $1\leq i\leq j\leq n$, by $T[i..j]$.
An {\em interval} of $T$ is any set $I$ of  integers from $\Sigma$ such that there exist $i,j$ with $1\leq i\leq j\leq n$
 and $I=Set(T[i..j])$. Then $[i,j]$ is called a {\em location} of $I$ on $T$. A {\em maximal location} of $I$
on $T$ is any location $[i,j]$ such that neither $[i-1,j]$ nor $[i,j+1]$ is a 
location of $I$. 

When $T$ is the identity permutation $\Id_p : = (1\, 2\, \ldots\, p)$, we denote 
$(i..j):=\{i,$ $i+1, \ldots, j\}$, which is also $Set(\Id_p[i..j])$. Note that all intervals of $\Id_p$ are of this form,
and that each interval has a unique location on $\Id_n$. When $T$ is an arbitrary permutation on $p$ elements (denoted $P$ in this case), 
we denote by $\Pm$ the function which associates with each element of $P$ its position in $P$.
For a subsequence $P[i..j]$ of $P$, we also say that it is {\it delimited} 
by its elements $p_i$ and $p_j$ located at positions $i$ and $j$. These elements are the {\it delimiters} of 
$P[i..j]$ (note the difference between delimiters, which
are {\it elements},  and their {\it positions}). 

We now define common intervals of two sequences $T$ and $S$ of respective sizes $n_1$ and $n_2$:

\bdefin   \cite{didier2003common,schmidt2004quadratic}
A {\em common interval} of two sequences $T$ and $S$ over $\Sigma$ is a set $I$ of integers that is an 
interval of both $T$ and $S$. A $(T,S)$-{\em maximal location of $I$} is  any pair 
$([i,j],[y,z])$ of maximal locations of $I$ on $T$ (this is $[i,j]$) and respectively on $S$ (this is $[y,z]$).
\edefin 

\bex
Let $T=1\,2\,5\,2\,1\,4\,3\,1\,2\,6\,5$ and $S=5\,6\,4\,2\,3\,4\,1\,5$. Then $\{1,2\}$ is an interval
of $T$ with locations $[1,2]$, $[4,5]$ and $[8,9]$, which are also maximal locations, but is not
a common interval of $T$ and $S$. An example of common interval is $\{1, 2, 3, 4\}$ which has
five locations on $T$, namely $[4,7], [4,8], [4,9], [5,9]$ and $[6,9]$, and two locations on $S$, namely
$[3,7]$ and $[4,7]$. However, there is only one maximal location on each of $T$ and $S$, so that
there is only one $(T,S)$-maximal location of $\{1,2,3, 4\}$, namely $([4,9],[3,7])$.
\eex

The problem we are concerned with is defined below. We assume, without loss of generality, that
both sequences contain all the elements of the alphabet, so that $n_1,n_2\geq p$.

\bigskip

\noindent{\sc $(T,S)$-Common Intervals Searching}

\noindent\begin{tabular}{ll}
\hspace*{-0.2cm}{\bf  Input:}& \hspace*{-0.4cm} \begin{minipage}[t]{12.7cm} Two sequences $T$ and $S$ of respective 
lengths $n_1$ and $n_2$ over an alphabet $\Sigma=\{1, 2, \ldots, p\}$. 
\end{minipage}\\
\hspace*{-0.2cm}{\bf Requires:}&\hspace*{-0.4cm} \begin{minipage}[t]{12.7cm} Find all $(T,S)$-maximal locations
of common intervals of  $T$ and $S$, without redondancy.
\end{minipage}

\end{tabular}
\bigskip

To address this problem, assume we add a new element $X$ (not in $\Sigma$) at positions 0 and $n_1+1$ of $T$. 
Let $\Succ$ be the  
$(n_1+1)$-size array defined for each position $i$ with $0\leq i\leq n_1$ by $\Succ[i]=j$ if  $t_i=t_j$ 
and $j>i$ is the smallest with this property  (if $j$ does not exist, then $\Succ[i]=n_1+1$). Call
the {\em area} of the position $i$ on $T$ the sequence $A_i: =T[i.. \Succ[i-1]-1]$.

\bex
With $T=X\,1\,2\,5\,2\,1\,4\,3\,1\,2\,6\,5\,X$, we have $\Succ=[12, 5, 4, 11, 9, 8, 12, 12, 12, 12, 12,$ $12]$.
Thus the area of position $2$ in $T$ is $A_2=T[2..\Succ[1]-1]=T[2..(5-1)]=2\,5\,2$.
Similarly, $A_4=T[4..\Succ[3]-1]=T[4..10]=2\,1\,4\,3\,1\,2\,6$.
\eex

\bdefin \cite{didier2003common}
The {\em order} $\Od_i$ associated with a position $i$ of $T$, $1\leq i\leq n_1$, is the
sequence of all elements in $Set(A_i)$ ordered according to their first occurrence in $A_i$.
We note $k_i=|Set(A_i)|=||\Od_i||$.  
\edefin

\br
Note that:

$\bullet$ $\Od_i$ may be empty, and this holds iff $t_{i-1}=t_i$. 

$\bullet$ if $\Od_i$ is not empty, then its first element is $t_i$.

$\bullet$ if $\Od_i$ is not empty, then $\Od_i$ contains each element in $Set(A_i)$ exactly once, 
and is thus a permutation on a subset of $\Sigma$. 
\er

In the subsequent, we consider that a pre-treatment has been performed on $T$,  removing every element $t_i$
which is equal to  $t_{i-1}$, $2\leq i\leq n_1$, such that to guarantee that no empty order exists. In
this way, the maximal locations are slightly modified, but this is not essential.

Let respectively $\bb_1(=i), \bb_2, \ldots, \bb_{k_i}$ be
the positions in $T$ of the elements $a_1(=t_i),a_2, \ldots, a_{k_i}$ defining $\Od_i$, i.e. the position in $T$ of their first occurrences in $A_i$.    Now, define $\B_i : =\bb_1\, \bb_2\, \ldots\, \bb_{k_i}$ to be the ordered
sequence of these positions.

\bex
With $T=1\,2\,5\,2\,1\,4\,3\,1\,2\,6\,5$, we have $A_4=2\,1\,4\,3\,1\,2\,6$ and thus
$\Od_4=2\,1\,4\,3\,6$ with $\B_4=4\,5\,6\,7\, 10$. Note that $\Od_4[1..4]=2\,1\,4\,3$
and $\B_4[1..4]=4\,5\,6\,7$ meaning that $Set(\Od_4[1..4])$, i.e. $\{1, 2, 3, 4\}$, is
an interval of $T$ a location of which is given by $\B_4[1]$ and $\B_4[4]$, i.e. $[4,7]$.
This location is not maximal, but  it is the {\em maxmin} location corresponding to
$[4,9]$ as defined below.
\eex 

\bdefin  Given a sequence $T$ and an interval $I$ of it, a {\em maxmin location} of $I$ on
$T$ is any location $[i,j]$ of $I$ which is left maximal and right minimal, that is, such that neither
$[i-1,j]$ nor $[i,j-1]$ is a location of $I$ on $T$.  A $(T,S)$-{\em maxmin location of $I$} is  any pair 
$([i,j],[y,z])$ of maxmin locations of $I$ on $T$ (this is $[i,j]$) and respectively on $S$ (this is $[y,z]$).
\edefin 

It is easy to see that that maxmin locations and  maximal locations are in bijection. We
make this more precise as follows.

\begin{fait} 
The function associating with each maximal location $[i,j']$ of an interval in $T$ the maxmin location $[i,j]$ 
in $T$ such  that $j$ is maximum with the properties $j\leq j'$ and $j\in Set(\B_i)$ is a bijection. Moreover, 
if $j=\B_i[h]$, then $j'$ may be computed in $O(1)$ when $i, h, \B_i$ and $Succ$ are known.
\label{claim:maxloc}
\end{fait}

{\bf Proof.} It is easy to see that by successively removing 
from $T[i..j']$ the rightmost element as long as it has a copy on its left, we obtain a unique interval 
$T[i..j]$ such that $[i,j]$ is a minmax location of $I$,  $j\in Set(\B_i)$ and $j$ is maximum with 
this property. The inverse operation builds $[i,j']$ when $[i,j]$ is given.

Moreover, if $j=\B_i[h]$, then $Set(T[i..j'])=Set(T[i..j])=Set(\Od_i[1..h])$. Then, assuming
$i,h, \B_i$ and $Succ$ are known and we want to compute $j'$, we have two cases. If $h=k_i$,
then $j'$ is the position of the last element in $A_i$ and thus $j'$ is computed as $j'=\Succ[i-1]-1$. If $h<k_i$, then
$j'$ is the position in $T$ of the element preceding $\Od_i[h+1]$, that is, $j'=\B_i[h+1]-1$.$\Box$ 
\bigskip

In the subsequent, and due to the preceding Claim, we solve the $(T,S)$-{\sc Common Interval Searching} problem by replacing
maximal locations with maxmin locations. Using Claim \ref{claim:maxloc}, it is also easy to deduce that:

\begin{fait} \cite{didier2003common}
The intervals of $T$ are the sets $Set (\Od_i[1..h])$ with $1\leq h\leq k_i$. As a consequence, the 
common intervals of $T$ and $S$ are the sets $Set(\Od_i[1..h])$ with $1\leq h\leq k_i$, which 
are also intervals of $S$. 

\label{claim:DidierCI}
\end{fait}

With these precisions, Didier's approach \cite{didier2003common}  consists then in considering each order $\Od_i$ and, in total time 
$O(n_2\log  n_2)$ (reducible to $O(n_2)$ according to \cite{schmidt2004quadratic}), verifying whether the intervals 
$Set(\Od_i[1..h])$ with $1\leq h\leq ||\Od_i||$ are also intervals of $S$. Our approach avoids to consider each
order $\Od_i$ by defining dominating orders which contain other orders, with the aim of focalising the search
for common intervals on each dominating order rather than spreading it on each of the orders it dominates.

We introduce now the supplementary notions needed by our algorithm. 

\bdefin
Let $d,i$ be two integers such that $1\leq d\leq i\leq n_1$. We say that the order $\Od_d$ {\em dominates} 
the order $\Od_{i}$ if $\B_i$ is a contiguous
subsequence of $\B_d$. We also say that $\Od_i$ is {\em dominated} by $\Od_d$.
\edefin

Equivalently, $\Od_i$ is a contiguous subsequence of $\Od_d$ and the positions on $T$ of their common elements are 
the same.

\bdefin
Let $d$ be such that $1\leq d\leq n_1$. Order $\Od_d$ is {\em dominating} if it is not dominated by any other order of $T$. 
The number of dominating orders of $T$ is the {\em domination number} $q(T)$ of $T$.
\edefin

The set of orders of $T$ is provided with an order, defined as $\Od_i\prec \Od_j$ iff  $i<j$.
For each dominating order $\Od_d$ of $T$, its {\em strictly dominated orders} are the
orders $\Od_i$ with $i\geq d$ such that $\Od_i$ is dominated by $\Od_d$ but is not dominated by any
order preceding $\Od_d$ according to $\prec$. 

\begin{figure}[t]
\vspace*{-2cm}
\begin{center}
\hspace*{-1cm}\includegraphics[width=12cm]{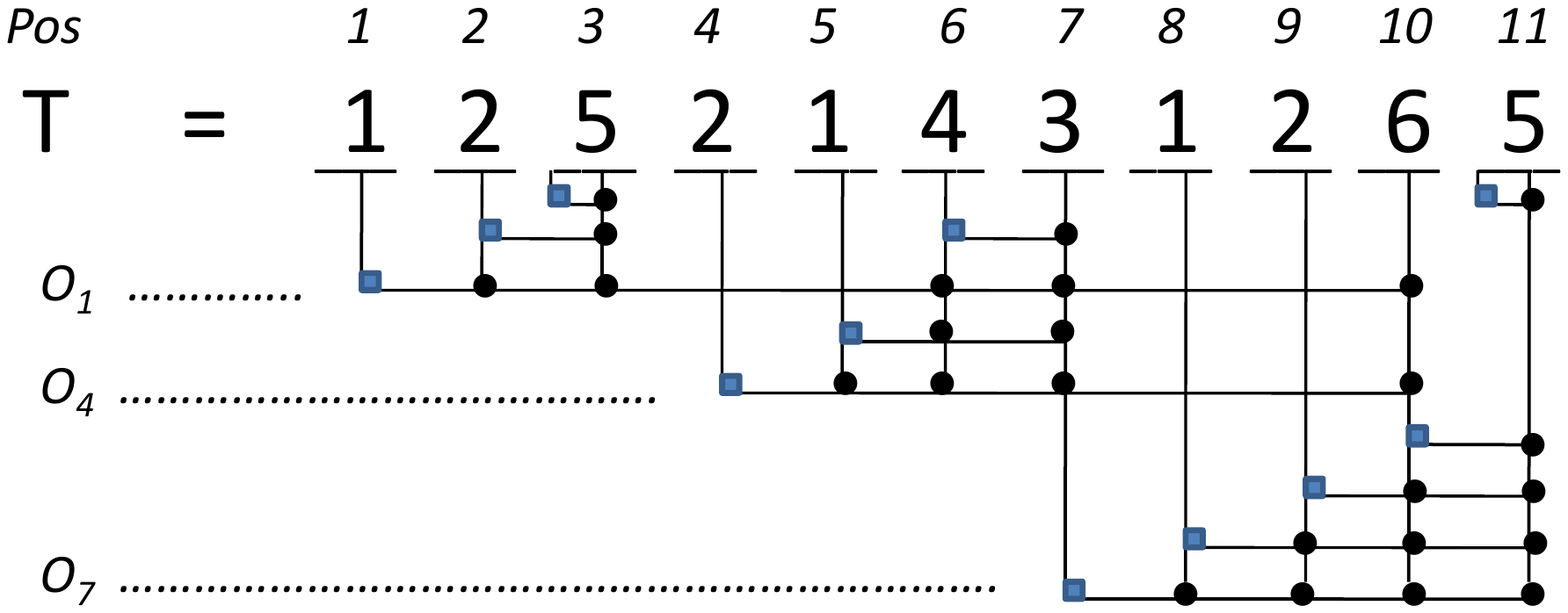}
\end{center}
\vspace*{-5cm}
\caption{\small The orders of $T=1\,2\,5\,2\,1\,4\,3\,1\,2\,6\,5$. For a given position $i$ of $T$, the order
$\Od_i$ is represented by the horizontal line whose intersection with the vertical line going down
from $t_i$ is marked with a square. The elements of $\Od_i$ are $t_i$ (marked with the abovementioned
square) and all the elements on the line $\Od_i$ marked with a circle. When an order contains only
one element, as $\Od_3$ and $\Od_{11}$, both the square and the circle represent the unique element
of the order. }
\label{fig:ordresT}
\end{figure}

\bex
The orders of $T=1\,2\,5\,2\,1\,4\,3\,1\,2\,6\,5$ are given in Figure \ref{fig:ordresT}. Orders
$\Od_1, \Od_4$ and $\Od_7$ are dominating. Note that $\Od_1$ strictly dominates $\Od_1, \Od_2, \Od_3$ and
$\Od_6$ although $\Od_6$ is also dominated by $\Od_4$. Similarly, $\Od_4$ strictly dominates $\Od_4$ and 
$\Od_5$. Finally, $\Od_7$ strictly dominates $\Od_7, \Od_8, \Od_9, \Od_{10}$ and $\Od_{11}$. The values of 
each sequence $\B_i$ are obtained by recording the positions instead of the values when considering a horizontal line.

\eex

For each dominating order $\Od_d$ (which is a permutation), we need to record the suborders which correspond
to the strictly dominated orders. Only the left and right endpoints of each suborder are recorded, in
order to limit the space and time requirements. Then, let the {\em domination function}  of a dominating order $\Od_d$ 
be the partial function $F_d:\{1, 2, \ldots, k_d\}\rightarrow \{1, 2, \ldots, k_d\}$ defined as follows.

$$F_d(s):=  f\,\,  \mbox{if}\,\, \mbox{there is some}\, i\, \mbox{such that}\, \Od_i\, \mbox{is strictly dominated by}\, \Od_d\, \mbox{and}\, \B_d[s..f]=\B_i.$$

\begin{figure}[t]
\vspace*{-2cm}
\begin{center}
\hspace*{-1cm}\includegraphics[width=12cm]{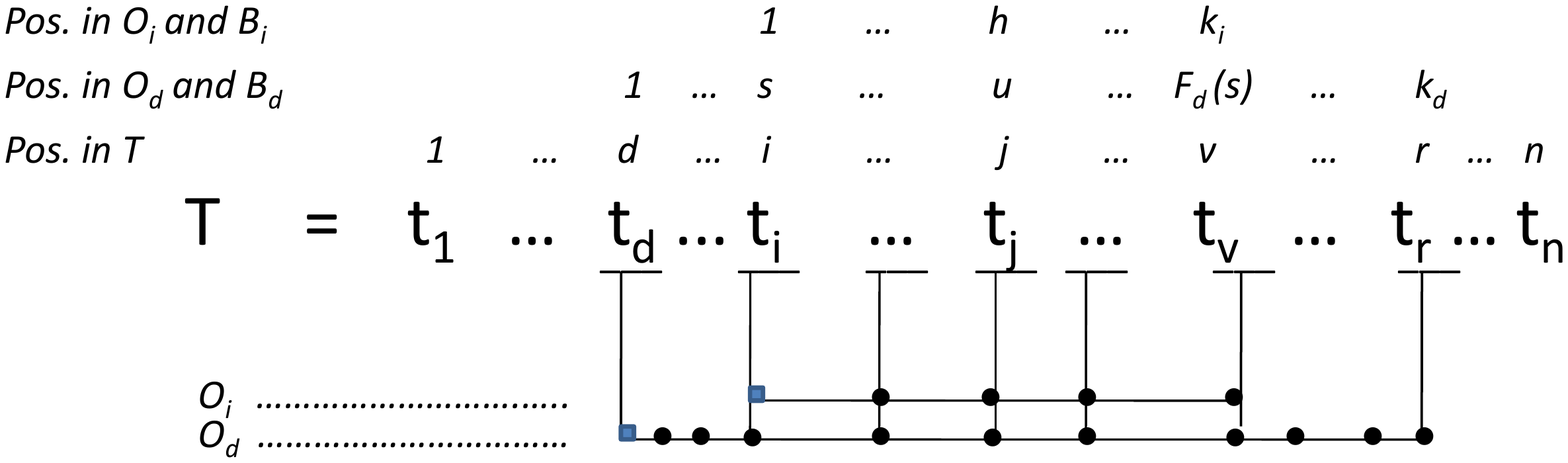}
\end{center}
\vspace*{-4cm}
\caption{\small Correspondence of positions between $T$, a dominating order $\Od_d$ of $T$ and an 
order $\Od_i$ of $T$ which is dominated by $\Od_d$. Black circles in $\Od_d$ and $\Od_i$ not identified by a position are other elements of 
$\Od_d$ and $\Od_i$, not important here.}
\label{fig:positionsT}
\end{figure}

\noindent For the other values of $s\in  \{1, 2, \ldots, k_d\}$, $F_d(s)$ is not defined. Note that
$F_d(1)=k_d$, since by definition any dominating order strictly dominates itself. See Figure \ref{fig:positionsT}.

\bex
For $T=1\,2\,5\,2\,1\,4\,3\,1\,2\,6\,5$ (see also Figure \ref{fig:ordresT}), the dominating order
$\Od_4$ strictly dominates $\Od_4$ and $\Od_5$, which correspond respectively to the
suborders $\Od_4[1..6]$ and $\Od_4[2..4]$ of $\Od_4$. The dominating function
of $\Od_4$ is then given by $F_4(1)=6$ and $F_4(2)=4$ ($F_4$ is not defined for the other values).  
\eex

We know that, according to Claim \ref{claim:DidierCI}, the common intervals of $T$ and $S$
must be searched among the intervals $Set(\Od_i[1..h])$ or, if we focus on one 
dominating order $\Od_d$ and its strictly dominated orders identified by $F_d$, among the 
intervals $Set(\Od_i[s..u])$ for which $F_d(s)$ is defined and  $s\leq u\leq F_d(s)$. We formalize this search as follows.

\bdefin Let $P$ be a permutation on $p$ elements, and $F:\{1, 2, \ldots, p\}\rightarrow \{1, 2, \ldots, p\}$  
be a partial function such that $F(1)=p$ and  $w\leq F(w)$ for all values $w$ for which $F(w)$ is defined.
A location $[s,u]$ of an interval of $P$  is {\em valid} with respect to $F$ if 
$F$ is defined for $s$ and $s\leq u\leq F(s)$.
\edefin

\begin{fait}
The $(T,S)$-maxmin locations $([i,j],[y,z])$ of  common intervals of $T$ and $S$ are in bijection with
the triples $(d, [s,u],[y,z])$ such that:

$(a)$ $\Od_d$ is a dominating order of\, $T$

$(b)$ the location $[s,u]$ on $\Od_d$ of the interval $Set(\Od_d[s..u])$ is valid with respect to $F_d$

$(c)$ $[y,z]$ is a maxmin location of $Set(\Od_d[s..u])$ on $S$.

Moreover, the triple associated with $([i,j],[y,z])$ satisfies : $\Od_d$ is the dominating order that strictly 
dominates $\Od_i$, $i=\B_d[s]$ and $j=\B_d[u]$.
\label{claim:couple}
\end{fait}

{\bf Proof.} See Figure \ref{fig:positionsT}. By Claim \ref{claim:DidierCI}, the common intervals of $T$ and $S$ are the
sets $Set(\Od_i[1..h])$ with $1\leq h\leq k_i$ which are intervals of $S$.  We note that
the sets $Set(\Od_i[1..h])$ are not necessarily distinct, but their locations $[i,j]$ on $T$,
given by $[i,j]=[\B_i[1],\B_i[h]]$, are distinct. Then, the $(T,S)$-maxmin locations $([i,j],[y,z])$ of 
common intervals $Set(\Od_i[1..h])$ are in bijection with the pairs $(\B_i[1..h], [y,z])$ such that
$[y,z]$ is a maxmin location of the interval on $S$, which are themselves in
bijection with the pairs $(\B_d[s..u], [y,z])$ such that the dominating order $\Od_d$ strictly 
dominates $\Od_i$ and $[s,u]$ is valid with respect to $F_d$. More precisely, $u=s+h-1\leq F_d(c)$. $\Box$

\bcor
Each $(T,S)$-maxmin location $([i,j],[y,z])$ of a common interval of\, $T$ and $S$ is computable in $O(1)$ time
if the corresponding triple $(d,[s,u],[y,z])$ and the sequence $\B_d$ are known.
\label{cor:maxmin}
\ecor

Looking for the $(T,S)$-maxmin locations of the common intervals of $T$ and $S$ thus reduces to 
finding   the $(\Od_d,S)$-maxmin locations of common intervals for each dominating order $\Od_d$ and for $S$, whose 
locations on $\Od_d	$  are valid with respect to the dominating function $F_d$ of $\Od_d$. The central problem to solve 
now is thus the following one (replace $\Od_d$ by $P$, $F_d$ by $F$  and $k_d$ by $p$):
\bigskip

\noindent{\sc $(P,S)$-Guided Common Intervals Searching}

\noindent\begin{tabular}{ll}
\hspace*{-0.2cm}{\bf  Input:}& \hspace*{-0.3cm} \begin{minipage}[t]{12.7cm} A permutation $P$ on $p$ elements, a sequence $S$ of length $n_2$ on the same set of $p$ elements, a partial function $F:\{1, 2, \ldots, p\}\rightarrow \{1, 2, \ldots, p\}$ $ $ such that $F(1)=p$ and  $w\leq F(w)$ for all $w$ such that $F(w)$ is defined.\smallskip\end{minipage}\\
\hspace*{-0.2cm}{\bf Requires:}&\hspace*{-0.3cm} \begin{minipage}[t]{12.7cm} Find all $(P,S)$- maxmin locations of common intervals of $P$ and $S$ whose locations on $P$ are valid with respect to $F$, without redondancy.
\end{minipage}

\end{tabular}
\bigskip

As before, we assume w.l.o.g. that $S$ contains all the elements in $P$, so that $n_2\geq p$. Also, we denote
$q_2:=q(S)$. In this paper,  we show (see Section \ref{sect:approach}, Theorem \ref{thm:guided}) that {\sc $(P,S)$-Guided 
Common Intervals Searching} may be solved in $O(q_2n_2+N_{P,S})$ time and $O(n_2)$ space, where $N_{P,S}$ 
is its number of solutions for $P$ and $S$. This running time gives the running time of our general algorithm.
However, an improved running time of  $O(n_2+N_{P,S})$ for solving {\sc $(P,S)$-Guided 
Common Intervals Searching} would lead to a $O(q_1n_1+q_1n_2+N)$ algorithm for the case of two sequences,
improving the complexity of the existing $O(n_1n_2)$ algorithms.

\section{The approach} \label{sect:approach}

The main steps for finding the maxmin locations of all common intervals in two sequences
using the reduction to  {\sc $(P,S)$-Guided Common Intervals Searching}  are given in Algorithm \ref{algo:main}.
Recall that for $T$ and $S$ we respectively denote $n_1, n_2$ their sizes, and $q_1, q_2$ their dominating numbers.
The algorithms for computing each step are provided in the next sections. 

To make things clear, we note that the dominating orders (steps 1 and 2) are computed but
never stored simultaneously, whereas dominated orders are only recorded as parts of their corresponding
dominating orders, using the domination functions. The initial algorithm for computing this information,
in step 1 (and similarly in step 2), is too time consumming to be reused in steps 3 and 4 when dominating 
orders are needed. Instead, minimal information from steps 1 and 2 is stored, which allows to recover 
in steps 3 and 4 the dominating orders, with a more efficient algorithm. In such a way, we keep the
space requirements in $O(n_1+n_2)$, and we perform steps 3, 4, 5 in global time $O(q_1q_2p)$, which
is the best we may hope. 

\begin{algorithm}[t,boxed]
\caption{Main algorithm} 
\begin{algorithmic}[1]
\REQUIRE Sequences $T$ and $S$ over the alphabet $\Sigma=\{1, 2, \ldots, p\}$.
\ENSURE  All $(T,S)$-maxmin locations of  common intervals of $T$ and $S$, without redondancy.

\STATE Compute the $q_1$ dominating orders of $T$ and their dominating functions $F_d$, $d=1, 2,\ldots, q_1$
\STATE Compute the $q_2$ dominating orders of $S$ and their dominating functions $\Phi_\delta$, $\delta=1, 2, \ldots, q_2$ 
\FOR{each dominating order $\Od_d$ of $T$}
\FOR{each dominating order $\Omega _{\delta}$ of $S$}
\STATE Compute the common intervals of $\Od_d$ and $\Omega_{\delta}$ that are valid w.r.t. $F_d$ and $\Phi_{\delta}$
\ENDFOR
\ENDFOR
\end{algorithmic}
\label{algo:main}
\end{algorithm}

In order to solve  {\sc $(P,S)$-Guided Common Intervals Searching},
our algorithm cuts $S$ into dominating orders and then it looks for common intervals in
permutations. This is done in steps 2, 4 and 5, as proved in the next theorem.

\begin{thm}
Steps 2, 4 and 5 in Algorithm \ref{algo:main} solve {\sc $(P,S)$-Guided Common Intervals Searching}
with input $P=\Od_d$, $F=F_d$ and $S$. Moreover, these steps may be performed in global $O(q_2n_2+N_{P,S})$ 
time and $O(n_2)$ space.
\label{thm:guided}
\end{thm}

{\bf Proof.} Claim \ref{claim:couple} and Corollary \ref{cor:maxmin} insure that the $(S,\Od_d)$-maxmin locations of common intervals
of $S$ and $\Od_d$, in this precise order, are in bijection with (and may be easily computed from) the triples $(\delta,[s,u],[y,z])$
such that $\Omega_\delta$ is a dominating order of $S$, $[s,u]$ is valid with respect to $\Phi_\delta$ 
and $[y,z]$ is a maxmin location of $Set(\Omega_\delta[s..u])$ on $\Od_d$. Note that since
$\Od_d$ is a permutation, each location is a maxmin location. Reducing these triples to those for which
$[y,z]$ is valid w.r.t. $F_d$, as indicated in step 5, we obtain the solutions of {\sc $(P,S)$-Guided Common Intervals Searching}
with input $P=\Od_d$, $F=F_d$ and $S$.

In order to give estimations of the running time and memory space, we refer to results proved in the
remaining of this paper. Step 2 takes $O(q_2n_2)$ time and $O(n_2)$ space
assuming the orders are not stored (as proved in Section \ref{sect:find}, Theorem \ref{thm:store}),
step 4 needs $O(q_2p)$ time and $O(n_2)$ space to successively generate the orders $\Omega_\delta$
from information provided by step 2 (Section \ref{sect:retrieve}, Theorem \ref{thm:retrieve}),
whereas step 5 takes $O(p+N_{\Od_d, \Omega_\delta})$ time and $O(p)$ space, where $N_{\Od_d, \Omega_\delta}$
is the number of solutions for  {\sc $(\Od_d,\Omega_\delta)$-Guided Common Intervals Searching}
(Section \ref{sect:common}, Theorem \ref{thm:common}).
 $\Box$

\bex
With $T=1\, 2\, 5\, 2\, 1\, 4\, 3\,1\, 2\, 6\, 5$ and $S= 5\, 6\, 4\, 2\, 3\, 4\, 1\, 5$
we have three dominating orders in $T$, that is $\Od_1, \Od_4$ and $\Od_7$, and three
dominating orders in $S$, that is $\Omega_1, \Omega_3$ and $\Omega_5$. Consider step 5 for
$\Od_4=2\,1\,4\,3\,6$ and $\Omega_5=3\, 4\, 1\, 5$. We have $F_4(1)=6$  and $F_4(2)=4$, as
well as $\Phi_5(1)=4$ and $\Phi_5(2)=4$ (note that  $\Phi_5(3)$ and $\Phi_5(4)$ are not defined as $\Omega_7$
and $\Omega_8$ are strictly dominated by $\Omega_3$ and not by $\Omega_5$). That means
we only look for common intervals which start in positions $1$ or $2$ in
$\Od_4$ and in positions $1,$ or $2$  in $\Omega_5$. Moreover, an interval which starts
in position $s$ must end not later than $F_4(s)$ in $\Od_4$ (and similarly for $\Omega_5$).
Thus the common intervals the algorithm will find for those two permutations are
$\{1,4\}$ (with locations $[2,3]$ in $\Od_4$, and $[2,3]$ in $\Omega_5$) and
$\{1, 3, 4\}$ (with locations $[2,4]$ in $\Od_4$ and $[1,3]$ in $\Omega_5$). 
Note that these locations are valid with respect to $F_4$ and $\Phi_5$. The
common interval $\{3,4\}$ of $\Od_4$ and $\Omega_5$ is not output in this step of the algorithm
since its location in $\Od_4$ is not valid. However, this is not a loss  since such an interval
would be redundant with the one output when $\Od_1$ and $\Omega_5$ are compared.
Also note that $\Od_4$ and $\Omega_5$ are not permutations on the same set, and thus the algorithm 
we give in Section \ref{sect:common} must be applied on two slightly modified permutations.
\eex

\begin{thm}
Algorithm \ref{algo:main} solves the {\sc $(T,S)$-Common Intervals Searching} problem in 
$O(q_1n_1+q_2n_2+q_1q_2p+N)$ time, where $N$ is the size of the solution,
and $O(n_1+n_2)$ space. 
 
\end{thm}

{\bf Proof.} The correctness of the algorithm is insured by Claim \ref{claim:couple}
and Theorem \ref{thm:guided}.

We now discuss the running time and memory space, once again referring to results
proved in the remaining sections. As proved in Theorem \ref{thm:store}
(Section \ref{sect:find}), Step 1 (and similarly Step 2) takes  $O(q_1n_1)$-time 
and $O(n_1)$ space, assuming that the dominating orders $\Od_d$
are identified by their position $d$ on $T$ and are not stored (each of them is computed, used to
find its dominating function and then discarded). The positions $d$ corresponding to dominating orders
are stored in decreasing order in a stack $D_T$. The values of the dominating functions are stored
as $q_1$ lists, one for each dominating order $\Od_d$, whose elements are the pairs $(s, F_d(s))$,
in decreasing order of the value $s$.  This representation needs a global memory space of $O(n_1)$.

In step 3 the progressive computation of the $q_1$ dominating orders is done in $O(q_1p)$ time and $O(n_1)$ space
using the sequence $T$ and the list $D_T$ of positions $d$ of the dominating orders. The algorithm achieving 
this is presented in Section \ref{sect:retrieve}, Theorem \ref{thm:retrieve}. For each dominating order 
$\Od_d$ of $T$, the orders $\Omega_\delta$ of $S$  are successively computed in global $O(q_2p)$ time and $O(n_2)$ 
space by the same algorithm, and are only temporarily stored. Step 5 is performed for $\Od_d$ and $\Omega_\delta$ 
in  $O(p+N_{\Od_d, \Omega_\delta})$ time and $O(p)$ space, where $N_{\Od_d, \Omega_\delta}$
is the number of output solutions for  {\sc $(\Od_d,\Omega_\delta)$-Guided Common Intervals Searching}
(Section \ref{sect:common}, Theorem \ref{thm:common}). 

Then the abovementioned running time of our algorithm easily follows. $\Box$
\bigskip

To simplify the notations, in the next sections the size of $T$ is denoted by $n$ and its
domination number is denoted $q$. The vector $\Succ$, as well as the vectors $\Precc$ and $\Precc_S$
defined similarly later, are assumed to be computed once at the beginning of Algorithm \ref{algo:main}.

\section{Finding the dominating and dominated orders of $T$} \label{sect:find}

This task is subdivided into two parts. First, the dominating orders $\Od_d$ are
found as well as, for each of them, the set of positions $i$ such that $\Od_d$
strictly dominates $\Od_i$. Thus $\Od_i=\Od_d[s..F_d(s)]$, where $s$ is
known but $F_d(s)$ is not known yet. In the second part of this section, we
compute $F_d(s)$. Note that in this way we never store any dominated order, but only 
its position on $T$ and on the dominating order strictly dominating it.
This is sufficient to retrieve it from $T$ when needed.

\subsection{Find the positions $i$ such that $\Od_i$ is dominating/dominated}\label{sect:findpos}

As before, let $T$ be the first sequence, with an additional element $X$ (new character) 
at positions 0 and $n+1$. Recall that we assumed that neighboring elements in $T$ are not equal,
and that we defined $\Succ$ to be the $(n+1)$-size array such that, for all $i$ with $0\leq i\leq n$,
$\Succ[i]=j$ if $t_i=t_j$ and $j>i$ is the  smallest with this property (if $j$ does not exist, then
$\Succ[i]=n+1$).

Given a subsequence $A=T[i..j]$ of $T$, {\em slicing} it into singletons means
adding the character $Y$ at the beginning and the end of $A$, as well as a so-called $h$-separator 
(denoted $|_h$) after each element of $A$ which is the  letter $h$. And this, for each $h$. 
Call $\As$ the resulting sequence on $\Sigma\cup\{Y\}\cup\{|_h\, |\, h\in \Sigma\cup\{Y\}\}$.

\bex
With $T=1\,2\,5\,2\,1\,4\,3\,1\,2\,6\,5$, let for instance $A: =T[4..10]= A_4= 2\,1\,4\,3\,1\,2\,6$. Slicing
$A$ into singleton yields $\As=Y\,|_Y\, 2\,|_2\,1\,|_1\,4\,|_4\,3\,|_3\,1\,|_1\,2\,|_2\,6\,|_6\,Y\,|_Y$.
\eex

Once $\As$ is obtained from $A$, successive {\it removals} of the
separators are performed, and the resulting sequence is still called $\As$. 
Let a  {\it slice} of $\As$ be any maximal interval $\{r, r+1, \ldots, s\}$ of positions in $\{i, \ldots, j\}$ 
(recall that $A=T[i..j]$) such
that no separator exists in $\As$ between $t_l$ and $t_{l+1}$ with $r\leq l<s$. 
Note that in this case a $t_{r-1}$-separator exists after $t_{r-1}$ and a $t_s$-separator exists 
after $t_s$, because of the maximality of the interval $\{r, r+1, \ldots, s\}$.
With $\As$ as defined above, immediately after $A$ has been sliced, every position in $A$ forms a slice.

\bex
With $A=T[4..10]$ and $\As$ obtained by slicing $A$ into singletons as in the preceding example, let
now $\As=Y\,|_Y\, 2\,|_2\,1\, 4\,|_4\,3\,|_3\,1\,2\,|_2\,6\,|_6\,Y\,|_Y$
be obtained after the removal
of all the $1$-separators.  The slices are now $\{4\}$ (corresponding to $t_4=2$), $\{5, 6\}$ (corresponding
to $t_5=1$ and $t_6=4$), $\{7\}$ (corresponding to $t_7=3$), $\{8,9\}$ (corresponding
to $t_8=1$ and $t_9=2$), and $\{10\}$ (corresponding to $t_{10}=6$).
\eex

Slices are disjoint sets which evolve from singletons to larger and larger disjoint
intervals using separator removals. Two operations are needed, defining - as the reader will
easily note - a Union-Find structure:

\begin{itemize}
\item[$\bullet$] Remove a $h$-separator, thus merging two neighboring
slices into a new slice. This is set union, between sets representing neighboring intervals.
\item[$\bullet$] Find the slice a position belongs to. In the algorithm we propose, this function is
denoted by $Find$.
\end{itemize}

In the following, a position $d$ is {\it resolved} if its order $\Od_d$ has already been identified,
either as a dominating or as a dominated order. Now, by calling Resolve($T,d$)
in Algorithm \ref{algo:resolve} successively for all  $d=1, 2, \ldots, n$ (initially non-resolved), 
we find the dominating
orders $\Od_d$ of $T$ and, for each of them, the positions $i$ such that $\Od_i$ is strictly dominated by $\Od_d$.
Note that the rightmost position of each $\Od_i$ dominated by $\Od_d$ is computed by the procedure RightEnd($d$),
given in Section \ref{findright}.

\begin{algorithm}[t,boxed]
\caption{Resolve($T,d$)}
\begin{algorithmic}[1]
\IF{$d$ is not resolved}
\STATE $A_d\leftarrow T[d..\Succ[d-1]-1]$; find $\Od_d$ and $\B_d$ by successively considering all elements in $A_d$
\STATE label $d$ as resolved and $\Od_d$ as dominating
\STATE $\As\leftarrow$ slice into singletons the sequence $A$ defined as $A: =T[d-1..\Succ[d-1]]$
\FOR{each position $i\in Set(\B_d)$ in decreasing order}
\STATE remove all the $t_i$-separators from $\As$ using $\Succ$
\IF{$i$ is not already resolved and $Find(i)=Find(\Succ[i-1])$}
 \STATE label $i$ as resolved and $\Od_i$ as dominated by $\Od_d$
 \ENDIF
\ENDFOR
\STATE RightEnd($d$) \hfill {\sl // after this step, we discard $A_d$, $\Od_d$ and $\B_d$}
\ENDIF
\end{algorithmic}
\label{algo:resolve}
\end{algorithm}

\bex
With $T=1\, 2\, 5\, 2\, 1\, 4\, 3\, 1\, 2\, 6\, 5$ and $d=1$, step 2 computes $\Od_1=1\, 2\, 5\, 4\, 3\, 6$
and $\B_1=1\, 2\, 3\, 6\, 7\, 10$ by searching into $A_1$. Then $1$ is labeled as resolved, and $\Od_1$ as
dominating. Further, $\As=Y\,|_Y\, 1\,|_1\, 2\,|_2\, 5\,|_5\, 2\,|_2\, 1\,|_1\, 4\,|_4\, 3\,|_3\, 1\,|_1\, 2\,|_2\, 6\,|_6\, 5\, |_5\, Y\,|_Y$. Starting the {\bf for} loop in step 5, with $i=10$ we remove the $6$-separator and
the condition in step 7 is not fulfilled. The same holds with $i=7$ after the $3$-separator is removed.
With $i=6$, the 4-separator is removed and the condition in step 7 is verified, so that $\Od_6$ is
labeled as dominated by $\Od_1$. Similarly, $\Od_3$, $\Od_2$ are further labeled as dominated by $\Od_1$. 
\eex
To prove the correctness of our algorithm, we first need two results.

\bfn
Order $\Od_i$ with $i>d$ is dominated by order $\Od_d$ iff $i<\Succ[i-1]\leq\Succ[d-1]$ and $i\in Set(\B_d)$  and $Set(T[d..i-1])\cap Set(A_i)=\emptyset$.
\label{claim:dom}
\efn

{\bf Proof.} Notice that, by definition, the positions in $\B_i$ belong to $\{i, i+1, \ldots, \Succ[i-1]-1\}$. 

''$\Rightarrow$'': Properties $i<\Succ[d-1]$ and $i\in \B_d$ are deduced directly
from the definitions of an order and of order domination. If the condition $\Succ[d-1]\geq \Succ[i-1]$ is not true, then
$\Succ[d-1]$ belongs to $\Od_i$ but not to $\Od_d$ (again by the definition of an order), a contradiction.
Moreover, if, by contradiction, there is some 
$r\in  Set(T[d..i-1])\cap Set(A_i)$, occurring respectively in positions $a$ and $b$ (choose each
of them as small as possible with $a>d$ and $b>i$), then $a\in Set(\B_d)$ and $b\in Set(\B_i)-Set(\B_d)$, 
since only the first occurrence of $r$ is recorded in $\Od_d$. But 
then $Set(\B_i)\not\subseteq Set(\B_d)$ and thus $\Od_i$ is not dominated by $\Od_d$, a contradiction.

''$\Leftarrow$'': Let $j\in Set(\B_i)$. Then the first occurrence of the element $t_j$ in $A_i$ is,
by definition, at position $j$. Moreover, $t_j\not\in Set(T[d..i-1])$ by hypothesis and since $\Succ[d-1]\geq \Succ[i-1]$, 
we deduce that the first occurrence of the element $t_j$ in $A_d$ is at position $j$. Thus $j\in Set(\B_d)$.
It remains to show that $\B_i$ is contiguous inside $B_d$. This is easy, since any position in $\B_d$,
not in $\B_i$ but located between two elements of $B_i$ would imply the existence of an element whose 
first occurrence in $A_d$ belongs to $A_i$; this element would then belong to $\Od_i$, and its position
to $\B_i$, a contradiction.$\Box$

\bfn
Let $d<i<\Succ[d-1]$, and assume $\Od_d$ is dominating. 
Then  $\Od_i$ is labeled as ''dominated by $\Od_d$'' in Resolve($T,d$) 
iff $\Od_i$ is strictly dominated by $\Od_d$.
\label{claim:b}
\efn

{\bf Proof.} Note that $\Od_i$ may get a label during Resolve($T,d$) iff $d$ is not resolved at the 
beginning of the procedure, in which case steps 2-3 of Resolve($T,d$)
insure that $\Od_d$ is labeled as ''dominating''. By hypothesis, we assume this label is correct.
Now, $\Od_i$ is labeled as ''dominated by $\Od_d$''  iff 

$\bullet$ $i\in \B_d$ (step 5), and 

$\bullet$ in step 7 we have that $i$ is not already resoved, and $i,\Succ[i-1]$ 
are  in the same slice in the sequence $\As$ where  all the $t_j$-separators satisfying  $j\in Set(\B_d)$ and $j\geq i$ 
have been removed (step~6). 

The latter of the two conditions is equivalent to saying that $A_i$ contains only characters equal to $t_j$,
$j\in Set(\B_d)$ and $j\geq i$, that is, only characters whose first occurrence in $A_d$ belongs to 
$A_i$.  This is equivalent to $Set(T[d..i-1])\cap A_i=\emptyset$ (i.e. no
character in  $A_i$ appears before $i$) and $\Succ[i-1]\leq \Succ[d-1]$ (all characters
in  $A_i$ have a first occurrence not later than $\Succ[d-1]-1$). But then the three
conditions on the right hand of Claim \ref{claim:dom} are fulfilled, and this means
$\Od_i$ is dominated by $\Od_d$. Given that step 8 is executed only once for a given position $i$,
that is, when $i$ is labeled as resolved, the domination is strict.$\Box$
\medskip

Now, the correctness of our algorithm is given by the following claim.

\bfn
Assume temporarily that the procedure $RightEnd(d)$ is empty. Then calling  Resolve($T,d$) successively 
for $d=1, 2, \ldots, n$ correctly identifies the 
dominating orders $\Od_d$ and, for each of them, the positions $i$ such that $\Od_d$ strictly 
dominates $\Od_i$. This algorithm takes $O(qn)$ time and $O(n)$ space.
\label{claim:sdom}
\efn

{\bf Proof.} We prove by induction on $d$ that, 
at the end of the execution of Resolve($T,d)$, we have for all $i$ with $1\leq i< \Succ[d-1]$: 

$(a)$ $\Od_i$ is labeled as ''dominating'' iff $i\leq d$ and $\Od_i$ is dominating

$(b)$ $\Od_i$ is labeled as ''dominated by $\Od_{d'}$'' iff $d'\leq d$ and $\Od_{d'}$ is  dominating and 
$\Od_i$ is strictly dominated by $\Od_{d'}$.

Say that a position $d$ is {\it used} if $d$ is unresolved when Resolve($T,d$) is called. 
We consider two cases.

{\sl Case $d=1$.} The position $d$ is necessarily used (no position is resolved yet), thus $\Od_d$ is labeled
as ''dominating'' (step 3) and no other order will have this label during the execution of Resolve($T,1$).
Now, $\Od_d$ is really  dominating, as there is no $d'<i$, and property $(a)$ is
proved. To prove $(b)$, recalling that $1< i< \Succ[d-1]$ and $d=1$, we apply Claim \ref{claim:b}.  Note
that $i\neq d$ since in step 7 $d$ is already resolved.

{\sl Case $d>1$.} Assume by induction the affirmation we want to prove is true before the call of
Resolve($T,d$). If $d$ is not used, that means $d$ is already resolved when Resolve($T,d$) is called,
and nothing is done. Properties $(a)-(b)$ are already satisfied due to the position $d'$ such
that $\Od_d'$ dominates $\Od_d$.

Assume now that $d$ is used. Then $\Od_d$ is labeled ''dominating'' and we have to show that $\Od_d$ is
really dominating. If this was not the case, then $\Od_d$ would be strictly dominated by some $\Od_{d'}$ with
$d'<d$, and by the inductive hypothesis it would have been labeled as so (property $(b)$ for $d'$). 
But this contradicts the assumption that $d$ is unresolved at the beginning of Resolve($T,d$). 
We deduce that $(a)$ holds.
To prove property $(b)$, notice that it is necessarily true for $d'<d$ and the corresponding dominated
orders, by the inductive hypothesis and since Resolve($T,d$) does not relabel any labeled order. 
To finish the proof of $(b)$, we apply Claim \ref{claim:b}.
When all the elements in $T$ are resolved, the algorithm stops and affirmations $(a)$-$(b)$ for the largest 
used $d$  guarantee that the labels are correct. 

The memory needed is obviously in $O(n)$ since $A_d$, $\Od_d$, $\B_d$ obtained in step 2 are only
stored during the call of Resolve($T,d$).  Step 2 needs $O(||A_d||)$ time by considering all elements $t_j$ in $A_d$ from left to 
right, adding their positions in $\B_d$ and using $\Succ$ to mark (so as to avoid considering them) all the
other elements in $A_d$ with the same value. The marks are discarded when step 2 is finished.

The running time of steps 4-10 is given by the implementation of the abovementioned Union-Find structure,
in which the universe is the set of positions from $A$, and the sets are the slices. 
As these sets always contain consecutive elements, and set unions are performed between 
neighboring slices, we are in the particular case of the Union-Find structure proposed in \cite{itai2006linear}.
With this structure,  a sequence of (intermixed)
$u$ unions and $f$ finds on a universe of $u$ elements is performed in $O(u+f)$ time. Here,
each $t_j$-separator in $\As$ is considered, and removed, exactly once thus implying one union between slices for
each $j$. 
Step 7 requires two $Find$ calls, for each element in $\B_d$. Overall, the running time is linear
in the size of $A$, which is in $O(n)$.

Then the running time of Resolve($T,d$) is in $O(n)$, for each dominating order $\Od_d$.
When the algorithm stops, the $q$ dominating orders and the start positions of the orders they strictly dominate 
are found in $O(qn)$ time. $\Box$

\subsection{Find, for each $i$ such that $\Od_i$ is dominated, the rightmost element of $\Od_i$}\label{findright}

In the preceding section, we found the dominating orders, which are of the type 
$\Od_d= a_1\, a_2\, \ldots\, a_{k_d}$, with $a_1=t_d$, for some $d$, and whose corresponding position
sequence is $\B_d=\bb_1\, \bb_2\, \ldots\, \bb_{k_d}$ (these are positions from $T$). 
This was done in step 2 of the algorithm Resolve($T,d$). For each such dominating order,  the 
positions $i\in \B_d$ with $i>d$ and such that $\Od_i$ is strictly dominated by $\Od_d$ have been 
identified in step 8 of Resolve($T,d$). For each such position $i$, assuming $i=\bb_h$, 
we must find now the endpoint $\bb_{f(h)}$, with $1<h\leq f(h)<k_d$, such
that $\B_i=b_h\, b_{h+1}\, \ldots\, b_{f(h)})=\B_d[h..f(h)]$. Recall Figure \ref{fig:positionsT}.

Consider Algorithm RightEnd($d$) in Algorithm \ref{algo:last}, where $St$ is a stack containing 
pairs of integers of the form $(h, \Succ[\bb_h-1])$, where  $\Od_{\bb_h}$ is strictly dominated
by $\Od_d$ and $h>1$. This algorithm computes, for each such $b_h$, a value $last[h]$.

\begin{algorithm}[t,boxed]
\caption{RightEnd($d$)}
\begin{algorithmic}[1]

\STATE $St\leftarrow \emptyset$
\FOR{$g\leftarrow 2$ to $k_d$}
\IF{$St\neq\emptyset$}
\STATE $(h,succ)\leftarrow\tp(St)$ \hfill{\sl // do not remove $(h,succ)$ from $St$}
\WHILE{$St\neq\emptyset$ and $\bb_g>succ$}
\STATE $last[h]\leftarrow g-1$
\STATE pop($St$)
\STATE {\bf if} $St\neq\emptyset$ {\bf then} $(h,succ)\leftarrow\tp(St)$ {\bf endif}\hfill{\sl // do not remove $(h,succ)$ from $St$}
\ENDWHILE
\ENDIF
\IF{$\Od_{\bb_g}$ is strictly dominated by $\Od_d$}
\STATE push $(g, \Succ[\bb_g-1])$ on $St$
\ENDIF
\ENDFOR
\WHILE{$St\neq\emptyset$}
\STATE $(h,succ)\leftarrow\tp(St)$
\STATE $last[h]\leftarrow k_d$
\STATE pop($St$)
\ENDWHILE
\end{algorithmic}
\label{algo:last}
\end{algorithm}

In order to show that $\B_i$ is indeed equal to $\bb_h (=i)\, \bb_{h+1}\, \ldots\, \bb_{last[h]}$ (i.e. $f(h)=last[h]$), we note that $last[h]$
has a value only if $h$ is on $St$ (steps 6 and 17), that is, only if $\Od_{\bb_h}$, with $\bb_h=i$, 
is strictly dominated by $\Od_d$ (these are the only positions pushed on $St$, by step 12). Therefore, 
we already know that $\B_{\bb_h}$ is a contiguous
subsequence of $\B_d$ which starts at $\bb_h$ and which finishes at the largest position of $\B_d$
which is smaller than $\Succ[\bb_h-1]$ (by the definition of an order). 
Then we must show that $last[\bb_h]$ is this element. 

\bex
With $T=1\, 2\, 5\, 2\, 1\, 4\, 3\, 1\, 2\, 6\, 5$ and $d=1$, we have $\Od_d=1\, 2\, 5\, 4\, 3\, 6$ and
$\B_d=1\, 2\, 3\, 6\, 7\, 10$. With $g=2$ and $g=3$ respectively, the pairs $(2,5)$ and $(3,4)$ are pushed
in this order on $St$. With $g=4$, we first have $(h,succ)=(3,4)$ and $4<b_4$ (which is 6), thus
$last[3]\leftarrow 3$. Then $(3,4)$ is discarded from $St$ and we have $(h,succ)=(2,5)$ and we deduce
$last[2]\leftarrow 3$. The stack $St$ is now empty, and $(4,8)$ is pushed on it. With $g=5$ nothing
happens, whereas with $g=6$ the algorithm produces $last[4]\leftarrow 5$. The stack is empty and 
RightEnd$(1)$ is finished.
\eex

We first prove that:

\bfn
Let $\bb_h$ and $\bb_g$, with $1< h<g\leq k_d$, be two positions such that $\Od_{\bb_h}$ and $\Od_{\bb_g}$ are
dominated by $\Od_d$. Then exactly one of the following statements holds:

$(a)$ $\Od_{\bb_h}$ dominates $\Od_{\bb_g}$

$(b)$ $A_{\bb_h}$ and $A_{\bb_g}$ are disjoint subsequences of $T$ such that $\bb_g> \Succ[\bb_h-1]$. Consequently, $\Od_{\bb_h}$ and $\Od_{\bb_g}$ are disjoint subsequences of $\Od_d$.
\label{claim:disjoint}
\efn 

{\bf Proof.} Obviously, the two affirmations cannot hold simultaneously. Now, assume 
by contradiction that none of them is true. Then  $\bb_g\leq \Succ[\bb_h-1]$. 
The value $t_{\bb_h-1}$ will then occur in $\Od_d$ before $\bb_h$
(since $d<\bb_h$), and will also occur in $\Od_{\bb_g}$, since $T[\Succ[\bb_h-1]]=t_{\bb_h-1}$. 
But then one cannot have $Set(T[d..\bb_g-1])\cap A_{\bb_g}=\emptyset$ and Claim \ref{claim:dom} is
contradicted. $\Box$

\bfn
In Algorithm RightEnd($d$), a pair $(g, \Succ[\bb_g-1])$ is on $St$ above $(h, \Succ[\bb_h-1])$ iff 
$h<g$ and $\Succ[\bb_g-1]\leq\Succ[\bb_h-1].$
\label{claim:stack}
\efn

{\bf Proof.}  We show this is true in the case where $(\bb_h, \Succ[\bb_h-1])$ is on the top of $St$
when $(g, \Succ[\bb_g-1])$ is pushed on $St$. The claim thus follows by induction. 

Assume then that $(\bb_h, \Succ[\bb_h-1])$ is on the top of $St$. The pair $(g, \Succ[\bb_g-1])$ is pushed on 
$St$ iff $\Od_{\bb_g}$
is strictly dominated by $\Od_d$ (steps 11-12), and $\bb_g\leq \Succ[\bb_h-1]$ (step 5).
Now, since $\bb_h<\bb_g$ according to step 2, we first deduce that property $(b)$ in Claim \ref{claim:disjoint}
does not hold.  Then property $(a)$ in Claim \ref{claim:disjoint} must be true. The claim follows
by Claim \ref{claim:dom}. $\Box$
\bigskip

Now, for each position $\bb_h$ such  that $\Od_{\bb_h}$ is strictly dominated by $\Od_d$, recall that 
$f(h)$ is such that $\B_{\bb_h}=\bb_h\, \ldots\, \bb_{f(h)}$. We prove that:

\bfn
At the end of Algorithm RightEnd($d$), we have  $last[h]=f(h)$ for each $h>d$ such that $\Od_{\bb_h}$ is strictly 
dominated by $\Od_d$. To achieve this, the algorithm takes $O(p)$ time and space.
\label{claim:last}
\efn

{\bf Proof.} It is easy to observe that we have $\bb_{f(h)}<\Succ[\bb_h-1]<\bb_{f(h)+1}$ (when $\bb_{f(h)+1}$ exists),
by the definition of an order and since $\B_{\bb_h}=\B_d[h..f(h)]$.

Let us follow the steps of Algorithm RighEnd($d$). When $g=h$ in step 2, $(h, \Succ[\bb_h-1])$ is pushed on $St$ in step 12. 
Since all the values $\bb_{h+1}, \ldots, \bb_{f(h)}$ are smaller than $\Succ[\bb_h-1]$, by the definition
of an order, when $g$ considers each of these values (step 2) the condition in step 5 is never fulfilled
with $succ=\Succ[\bb_h-1]$. Then $(h, \Succ[\bb_h-1])$ is still on the stack at the end of the execution of
the {\bf for} loop in step 2 for $g=f(h)$. Then we have two cases. 

\begin{itemize}
\item Either  $g\leftarrow f(h)+1$ is possible in step 2
(i.e. $f(h)<k_d$) and we deduce that  $\bb_g=\bb_{f(h)+1}>\Succ[\bb_h-1]$ by the observation above, 
and because of Claim \ref{claim:stack} we also have $\bb_{f(h)+1}>\Succ[\bb_{g'}-1]$ for all pairs $(g', \Succ[\bb_{g'}-1])$ which
are before $(h, \Succ[\bb_h-1])$ on $St$. The {\bf while} loop in steps 5-9 for $g=f(h)+1$ will thus discard all
these pairs, including $(h, \Succ[\bb_h-1])$. In step 6, we will then have $last[h]\leftarrow f(h)+1-1$,
i.e. $last[h]\leftarrow f(h)$.  
\item Or $f(h)=k_d$, and then in step 17 we have $last[h]\leftarrow k_d (=f(h))$.
\end{itemize}

In both cases, the proof of correctness is finished. Concerning the running time of
RightEnd($d$), its inner loop is the {\bf while} loop in step 5.
The other loops obviously have a total running time of $O(k_d)$, for a fixed $d$. The condition in
the {\bf while} loop is tested with an affirmative answer at most once for each pair
on the stack (and there are at most $k_d-1$ pairs), and with a negative answer  at most
once for each $g$. Thus, the overall execution of the algorithm, for a fixed $d$, is
in $O(k_d)$, and thus in $O(p)$. The space requirements are obviously in $O(p)$,
assuming the vector $\Succ$ is computed only once at the begining of the main algorithm.  $\Box$

\begin{thm}
Given a sequence $T$ of length $n$ over an alphabet $\Sigma$, there is an 
algorithm running in $O(qn)$ time and using $O(n)$ space to compute, without storing, the $q$ dominating orders of $T$ and,
for each such order $\Od_d$, its dominating function $F_d$. 
\label{thm:store}
\end{thm}

{\bf Proof.} By Claim \ref{claim:sdom}, Algorithm Resolve($T,d$) successively called for $d=1, 2, \ldots, n$ 
finds in $O(qn)$ time and $O(n)$ space the dominating orders $O_d$, and, for 
each order strictly dominated by $\Od_d$, the position $i$ on $T$ of its first element. 
Moreover, assuming that $\B_d=\bb_1\, \ldots\, \bb_{k_d}$, Algorithm RightEnd($d$) computes
for each $i(=\bb_h)$ the endpoint $f(h)$ with $1<h\leq f(h)\leq k_d$ such that
$\B_i=\B_d[h..f(h)]$. But then the pairs $(h,f(h))$, to which we must add the trivial pair $(1,k_d)$,
 are exactly the pairs $(s,F_d(s))$
in the definition of the dominating function $F_d$ of $\Od_d$:
\vspace*{-0.2cm}

$$F_d(s):=  f\,\,  \mbox{if}\,\, \mbox{there is some}\, i\, \mbox{such that}\,  \Od_i\, \mbox{is strictly dominated by}\, \Od_d\, \mbox{and}\, \B_d[s..f]=\B_i.$$

The running time of Resolve($T,d$) for all values of $d$ is in $O(qn)$, when the running time of the RightEnd($d$) call is left apart, 
as already proved  in Claim \ref{claim:sdom}.  For all $q$ dominating orders, RightEnd() takes $O(qp)$ time by
Claim \ref{claim:last}, and this does not change the overall 
running time. $\Box$

\section{Retrieving the dominating orders of $T$}\label{sect:retrieve}

Once the step 1 in Algorithm \ref{algo:main} is performed,  the list $D_T$ of
positions $d$ such that $\Od_d$ is a dominating order of $T$ is available. We assume it is a stack in
which the positions are ordered in decreasing order from top to bottom.  
Moreover, we assume a vector $\Precc$ has been built for $T$ (augmented
with character $X$ on both its endpoints) and is available, 
defined similarly to the vector $\Succ$. Vector $\Precc$ is the  
$(n+1)$-size array defined for each $i$ with $1\leq i\leq n+1$ by $\Precc[i]=j$ if  $t_i=t_j$ 
and $j<i$ is the largest with this property  (if $j$ does not exist, then $\Precc[i]=0$).
As was the case for $\Succ$, the array $\Precc$ can be built in $O(n)$ time.

Then we may retrieve all the dominating orders in global time of $O(qp)$ and using $O(n)$ space. More
precisely, we show that the dominating orders $\Od_d$, for all $d$, may be found in decreasing order
of $d$ in $O(qp)$ time. Note that, according to Algorithm \ref{algo:main}, we do
not need to store all the orders, but  only to progressively generate and  use them. Once used,
each order is discarded.

Consider Algorithm Retrieve($T, D_T,\Precc$) in Algorithm \ref{algo:retrieve}. The algorithm
works on a sequence $\W$ which is initially built such that $\W[i]=i$ and which reduces as the 
algorithm progresses. Intuitively, $\W$ stores the positions in $T$ of the elements of $T$ that
belong to at least one order remaining to be generated. Starting at the end 
of $\W$,  the algorithm 
finds in position $d$ of $\W$ the value $d$ that is on the top of $D_T$, i.e. the maximum
position of a dominating order.  Then the part of $\W$ between $d$ (which is the first
element of $\Od_d$) and a previously computed value $e$ (which turns out to be the last value in $\B_d$) 
is $\B_d$ and allows to easily deduce $\Od_d$. Finally, it remains to remove from $\W$ the positions in $T$
which became useless, and this is done by verifying, for each position in $\B_d$, if it is used or not by the
dominating order immediately following $\Od_d$ on its left, that is, $\Od_{d'}$ where $d'$ is the new
top of $D_T$.

\begin{algorithm}[t,boxed]
\caption{Retrieve($T, D_T, \Precc$)}
\begin{algorithmic}[1]

\STATE $w\leftarrow ||T||$; $W\leftarrow 1\, 2\, \ldots\, w$; $e\leftarrow w$
\WHILE{$D_T$ is not empty}
\STATE $d\leftarrow \tp(D_T)$; pop($D_{T}$)
\STATE $\B^*_d\leftarrow \W[d..e]$; $k^*_d\leftarrow e-d+1$; $\Od^*_d\leftarrow t_{\B^*_d[1]}\, t_{\B^*_d[2]}\, \ldots\, t_{\B^*_d[k^*_d]}$ \\  {\sl //here, $\Od^*_d$ and $\B^*_d$  may be used, and then discarded}
\IF{$D_{T}$ is not empty}
\STATE $j\leftarrow d$; $d\,'\leftarrow \tp(D_T)$
\WHILE{$j\leq w$}
\IF{$\Precc[\W[j]]=d\,'-1$ and $d\,'\neq 1$}
\STATE mark $j$ for removal from $\W$; $j\leftarrow w+1$
\ELSE
\STATE {\bf if} {$\Precc[\W[j]]\geq d\,'$} {\bf then} mark $j$ for removal from $\W$ {\bf else} $e\leftarrow j$ {\bf endif} 
\STATE  $j\leftarrow j+1$
\ENDIF
\ENDWHILE
\STATE remove all marked positions from $W$; update $e$ to point to the same value in $W$
\STATE $w\leftarrow ||W||$
\ENDIF
\ENDWHILE

\end{algorithmic}
\label{algo:retrieve}
\end{algorithm}

\bex
With $T=1\,2\,5\,2\,1\,4\,3\,1\,2\,6\,5$ and $D_T=\{7, 4, 1\}$ (7 on top, 1 on bottom), 
we have $\W=Id_{11}$ and in step 3 of 
Algorithm \ref{algo:retrieve} we have $d\leftarrow 7$ and $D_T\leftarrow\{4,1\}$, thus 
$\B^*_7\leftarrow W[7..11]= 7\, 8\,9\,10\,11$ resulting into $\Od^*_7=3\, 1\, 2\, 6\, 5$, which is
indeed $\Od_7$. In steps 9 and 11, we mark 11 and respectively 8,9 for removal. Values $8,9$ cannot
belong to another order, whereas 11 marks the end of the area $A_{4}$ for the top $d\,'=4$ of the 
stack $D_T$. Therefore, $W[e]$ points to the position in $T$ of the last validated element of $A_{4}$. 
This position is 10. After the removals, $W$ becomes $W= 1\,2\,3\,4\,5\,6\,7\,10$ and $e$ becomes 8
(so as $W[e]=10$ as before the removals). During the
new execution of the {\bf while} loop, with $d=4$, we have $\B^*_4=W[4..8]=4\,5\,6\,7\,10$
resulting into $\Od^*_4=2\,1\, 4\, 3\, 6$, which is $\Od_4$. And so on.
\eex

The proof makes use of the following easy result:

\begin{fait} Let $i, d, d'$ be positions of $T$ such that $d'<d<i$ and $i\in Set(\B_d)$.  If
$i\not\in Set(\B_{d'})$ then at least one of the following affirmations holds:

$(a)$ $i>Succ[d'-1]$ 

$(b)$ for all $d''\leq d'$, $i\not\in Set(\B_{d''})$.
\label{claim:before}
\end{fait}

{\bf Proof.} Assume affirmation $(a)$ does not hold. Then $i\not\in Set(\B_{d'})$ implies that either there is $k$ with 
$d'\leq k<d<i$ such that $t_{k}=t_i$, or $i=\Succ[d'-1]$, which may be written $t_{k}=t_i$ with $k=d'-1$. 
Then, for all $d''< d'$ 
we have $d''\leq k$ and thus $i$ cannot belong to $\B_{d''}$ since $t_i$ is preceded by $t_k$. $\Box$

\begin{fait}
After the execution of the {\bf while} loop in step 2 of Algorithm Retrieve($T, D_T, \Precc$) for\, $d\in D_T$, 
we have:

$(a)$ the values $\B^*_d$, $\Od^*_d$ and $k^*_d$ are equal respectively to $\B_d, \Od_d$ and $k_d$ 

$(b)$ $\B_{d'}=t_{W[d']}\,t_{W[d']+1}\,\ldots\, t_{W[e]}$

$(c)$  $\bigcup_{d''\in D_T, d''\leq d'} Set(\B_{d''})\subseteq Set(W)$. 
\label{claim:retrieve}
\end{fait}

{\bf Proof.} We use induction on the execution number $\alpha$ of the {\bf while} loop.

Case $\alpha=1$. Then $d$ is the first element in $D_T$, and thus $\Od_d$ is the dominating order of $T$ with
largest $d$. Moreover,  in $T[d..w]$ one cannot have $t_l=t_r$ for two positions $l,r$ (assume $l<r$) 
 since otherwise $\Od_r$
would be dominated neither by $\Od_d$ nor by an order preceding $\Od_d$ on its left, implying that
$d$ is not the top of $D_T$, a contradiction. Consequently, $\Od_d=T[d..w]$ and thus affirmation
$(a)$ holds, since $e=w$ by step 1. To see that affirmations $(b)-(c)$ hold, notice
that steps 8-9 identify and prepare for removal from $W$ the position $\Succ[d'-1]$ which marks the end 
of $A_{d'}$ in $T$, whereas step 11 prepares for removal from $W$ the positions $W[j]$ of $T$ which have
a copy on their left in $A_{d'}$. Affirmation $(b)$ follows by observing that $e$ keeps trace of the
last element which has no copy on its left in $A_{d'}$.
Also, by property $(b)$ in Claim \ref{claim:before}, we deduce affirmation $(c)$.

Case $\alpha>1$. Given that the new value of $d$ is the value of $d'$ during the $(\alpha-1)$-th
execution of the {\bf while} loop, the induction hypothesis (affirmation $(b)$) directly
implies affirmation $(a)$. To prove affirmation $(b)$, notice that the {\bf while} loop
in steps 7-14 considers all elements in $W$ between $d$ and the minimum between $\Succ[d'-1]$ and 
$w$ (the end of $W$). By the induction hypothesis (affirmation $(c)$) all positions in 
$\B_{d'}$ are in $W[d..w]$. As before, steps 8-9 and respectively 11 insure, also using Claim \ref{claim:before}, 
that the elements in $\B_{d'}$ are correctly selected (thus affirmation $(b)$ holds) and  
that the removed elements are now useless (thus affirmation $(c)$ holds).  $\Box$
 
\begin{thm}
Algorithm Retrieve($T, D_T, \Precc$) successively computes, without storing, the dominating orders of $T$ in
$O(pq)$ time and  $O(p)$ space. 
\label{thm:retrieve}
\end{thm}

{\bf Proof.} Affirmation $(a)$ in Claim \ref{claim:retrieve} guarantees the correctness of the algorithm. 
The running time of the algorithm assumes $\Precc$ has been computed once at the begining of the main algorithm.
Then the running time is given by the number of executions of the {\bf while} loop in steps 7-14
and by step 15. For each $d$, the {\bf while} loop in steps 7-14 considers all elements
in $W$ between $d$ and $\min\{\Succ[d'-1],w\}$. Overall, each position $j$ in $T$ is considered
at most once for each $d'$ such that $j\in Set(\B_{d'})$ (step 11) and exactly once for the largest
$d'$ such that $j\not\in \B_{d'}$ (step 9 or 11). Thus the overall running time of the
algorithm is in $O(pq)$ time, since the size of an order is at most $p$ and $pq\geq n$. 
Obviously, the only extra-space used is that for temporarily storing $\B_d$ and $\Od_d$ for
a fixed $d$. To finish the proof, note that step 15 may be performed in time proportional to
the number of elements removed from $W$ by representing $W$ as a double chained list. Two pointers
are then sufficient. One of them starts in $w$ (step~1) and goes back up to the element containing $d$,
which is also the $d$-th element of $W$. The other one is $j$, which goes from $d$ to $w$ in steps
4 and 7, and also allows to remove the marked elements in step 15. $\Box$

\br
Obviously, performing steps 3, 4 and 5 in the main algorithm (Algorithm \ref{algo:main}) requires to
call Retrieve($T, D_T,\Precc$), to insert a call of Retrieve($S, D_S, \Precc_S$) between steps 4 and 5 of 
Retrieve($T,$ $D_T,$ $\Precc$), and a call for the step 5 in the main algorithm between steps 4 and 5 of
Retrieve($S, D_S,\Precc_S$). Once a call from Retrieve() is finished, the information computed in
step 4 of Retrieve() is discarded.
\er

\section{Finding the common intervals of $\Od_d$ and $\Omega_{\delta}$}\label{sect:common}

Note here that $\Od_d$ and $\Omega_\delta$ are, by construction, permutations on no more than $p$
elements. Without loss of generality, we assume here that they are permutations on $\Sigma$ (otherwise
the same algorithm must be applied on two permutations obtained from the initial ones by adding
the missing elements at the end of each permutation, but {\it without} modifying the functions
$F_d, \Phi_\delta$).

Consequently, make the following changes of notation. Renumber the elements of $\Od_d$ such that $\Od_d$
becomes the identity permutation $\Id_p$ (simplified hereafter as $\Id$), and renumber the 
elements of $\Omega_\delta$ accordingly
so as to obtain a permutation $\pi$. Call $F$ the dominating function of $\Id$, and $\Phi$ that of $\pi$.

\br Note that the common intervals of $\Id$ and $\pi$ that are valid with
respect to $F$ and $\Phi$ are exactly the common intervals $(s..u)$ of $\Id$ and $\pi$ with
location $[y,z]$ on $\pi$ such that $F(s)$ and $\Phi(y)$ are defined, $u\leq F(s)$ and
$z\leq \Phi(y)$.
\er

Also notice that by Claim \ref{claim:disjoint} we have:

\br
Assume the dominating function $F_d$ of the dominating order $\Od_d$  is defined for  two values 
$s_0, s_1$ with $s_0< s_1$. Let $f_0=F_d(s_0)$ and $f_1=F_d(s_1)$. Then the orders
$\Od_d[s_0.. f_0]$ and $\Od_d[s_1..f_1]$  either dominate each other or are disjoint subsequences of $\Od_d$.
As a consequence, the similar affirmation holds for each of $F$ and $\Phi$.
\label{rmk:nested}
\er

In \cite{IR2013}, the $\LR$-Search algorithm in Algorithm \ref{algo:LR} is proposed for
finding the common intervals of $K$ permutations, for an arbitrary $K\geq 2$.
In order to show that this algorithm may be easily adapted to find common intervals
of $\Id$ and $\pi$ that are valid with respect to $F$ and $\Phi$, we need to present 
the details of the algorithm. Note that the algorithm looks for common intervals
of size of least 2 (those of size 1 are easy to obtain).

\subsection{The $\LR$-Search algorithm}

The presentation in this section follows very closely that in \cite{IR2013}. 

\begin{algorithm}[t,boxed]
\caption{The $\LR$-Search algorithm}
\begin{algorithmic}[1]
\REQUIRE Set $\PK$ of $K$ permutations over $\Sigma$, bounding functions $\BL$ and $\BR$,  \out$\,$ procedure
\ENSURE  All common intervals $(s..u)$ of $\PK$ with $u\in\Ra(s)$, filtered by \out

\STATE Compute $\mins^k,\mins$ and $\bs$ with $2\leq k\leq K$ and $s\in\{1, 2, \ldots, p\}$
\STATE Compute $\maxs^k,\maxs$ and $\Bs$ with $2\leq k\leq K$ and $s\in\{1, 2, \ldots, p\}$
\STATE Initialize an $\LmRp$-stack with empty stacks $L, R$
\FOR{$s\leftarrow p-1$ to $1$}
\STATE $\PopL(\bs)$ \hfill{\sl // discard from $L$ all candidates larger than $\bs$ and push $\bs$ instead}
\STATE $\PopR(\Bs)$ \hfill{\sl // discard from $R$ all candidates  smaller than $\Bs$}
\IF {$\Bs=s+1$}
\STATE $\PushLR(\bs,s+1)$ \hfill{\sl // $s+1$ is a new right candidate, suitable for each $s$ on $L$}
\ENDIF
\STATE Call \out$\,$ to choose a subset of intervals $(s..u)$ with $u\in \Ra(s)$
\ENDFOR
\end{algorithmic}
\label{algo:LR}
\end{algorithm}

Let $\PK=\{P_1, P_2, \ldots, P_K\}$ be a set of $K$ permutations over 
$\Sigma=\{1, 2, \ldots, p\}$ such that $P_1=\Id$. 
Now, let $\mins^k$ (respectively $\maxs^{\,k}$) 
be the minimum (respectively maximum) value in the interval of $P_k$ delimited by $s$ and $s+1$ (both included).
Also define 
$$\mins := \min\{\mins^k\,|\, 2\leq k\leq K\}, \maxs := \max\{\maxs^{\,k}\,|\, 2\leq k\leq K\}.$$
Note that, for each $k\in\{1, 2, \ldots, K\}$, $\mins\leq \mins^k\leq s < s+1\leq \maxs^{\,k}\leq \maxs$.

We call {\it bounding} functions $\BL,\BR:\Sigma \rightarrow \Sigma$ any two functions such that $\BL(s)\leq~\mins$ and $\BR(s)\geq \maxs$, for all $s\in \{1, 2, \ldots, p-1\}$.  We denote $\bs :=\BL(s)$ and $\Bs :=\BR(s)$ 
. 

\bdefin
Let $\PK$ be a set of permutations on $\Sigma$. Then the {\em $\M$-profile} of $\PK$ with respect to $\BL$ and $\BR$ 
is the set of pairs $[\bs,\Bs]$, $s\in\{1, 2, \ldots, p-1\}$. 
\edefin

The $\M$-profile of $\PK$ is the information needed by the $\LR$-Search algorithm (see Algorithm \ref{algo:LR})  to compute the common intervals
$(a..c)$ of $\PK$ containing both $\bs$ and $\Bs$ for all $s\in\{a, a+1, \ldots, c-1\}$. In this way, the
functions $\BL$ and $\BR$ allow a first selection among all common intervals in $\PK$. The \out\, procedure
in the input, used in step 10, completes the selection tools of the algorithm. During the computation, the 
interval candidates are stored
in an abstract data structure called an $\LR$-Stack.

\bdefin  \cite{IR2013}  An {\em $\LR$-stack} for an ordered set $\Sigma$ is a 5-tuple $(L, R, \SL, \SR, \Rt)$ such that:

\begin{itemize}
\item $L, R$ are stacks, each of them containing distinct elements from $\Sigma$ in either increasing or 
decreasing order (from top to bottom). The {\it first} element of a stack is its top, the {\it last} one is its bottom.
\item $\SL,\SR \subset \Sigma$ respectively represent the set of elements on $L$ and $R$.
\item $\Rt : \SL\rightarrow \SR$ is an injective function that associates with each $a$ from $\SL$ a pointer to an 
element on $R$ such that $\Rt(a)$ is before $\Rt(a')$ on $R$ iff $a$ is before $a'$ on $L$.
\end{itemize}

According to the increasing (notation +) or decreasing (notation -) order of the elements on
$L$ and $R$ from top to bottom, an $\LR$-stack may be of one of the four types $\LpRp,  \LmRm, \LpRm, \LmRp$.
\edefin 

\br We assume that each of the stacks $L,R$ admits the classical operations 
$pop, push$, and that their elements may be read without removing them. 
In particular, the function $\tp()$ returns the first element of the stack, without removing it,
and the function $\next(u)$ returns the element immediately following $u$ on the stack
containing $u$, if such an element exists. 
\label{rmk:next}
\er

  We further denote, for each $a\in\SL$ and with $a'=\next(a)$, assuming that $\next(a)$ exists:
$$\Ra(a)=\{c\in\SR\, |\, c\,\,  \mbox{\rm is located on}\,  R\,\, \mbox{\rm between}\, \Rt(a)\, \mbox{\rm included and}\,  \Rt(a')\, \mbox{\rm excluded}\}$$
When $\next(a)$ does not exist, $\Ra(a)$ contains all elements between $\Rt(a)$ included and the bottom of $R$ included.
Then $\Rt(a)$ is the first (i.e. closest to the top) element of $\Ra(a)$ on $R$.

We define the following operations on the LR-stack.
Note that they do not affect the properties of an $\LR$-stack. Sets $\Ra()$ are assumed to be updated without further specification whenever the pointers $\Rt()$ change.
Say that $a'$ is {\it $L$-blocking} for $a$, with $a'\neq a$, if $a$ cannot be pushed on $L$ when $a'$ is
already on $L$ (because of the increasing/decreasing order of elements on $L$),
and similarly for $R$.

\begin{itemize}
\item $\PopL(a)$, for  some $a\in \Sigma$: 
pop successively from $L$ all elements that are $L$-blocking for $a$, push $a$ on $L$
iff at least one $L$-blocking element has been found and $a$ is not already on $L$, 
and define $\Rt(\tp(L))$ as $\tp(R)$. At the end,
either $a$ is not on $L$ and no $L$-blocking element exists for $a$, or $a$ is on the top of $L$ and $\Rt(a)$ is a pointer to the top of $R$.

\item $\PopR(c)$, for some $c\in\Sigma$: pop successively from $R$ all elements that
are $R$-blocking for $c$, update all pointers $\Rt()$ (here, $\Rt(a)=nil$ is accepted temporarily if
$\Ra(a)=\emptyset$) and successively pop from $L$ all the elements $a$
with $\Rt(a)=nil$.  At the end, either $c$ is not on $R$ and no $R$-blocking element exists for $c$, or $c$ is on the top of $R$.

\item $\PushLR(a,c)$, for some $a,c\in \Sigma$ (performed when no $L$-blocking element exists for $a$ and
no $R$-blocking element exists for $c$): push $a$ on $L$ iff $a$ is not already on the top of $L$, 
push $c$ on $R$ iff $c$ is not already on the top of $R$,
and let $\Rt(\tp(L))$ be defined as $\tp(R)$.

\item $\FindL(c)$, for some $c\in\SR$: return the element $a$ of $\SL$ such that $c\in \Ra(a)$. (Note that
this operation is not explicitely used in the $\LR$-Search algorithm, but may be used in a separate
algorithm to solve its
steps 1 and 2 \cite{IR2013}).

\end{itemize}

\br
 Note that operations $\PopL(a)$ and $\PopR(c)$ perfom {\it similar} but not {\it identical} modifications
 on stacks $L$ and $R$ respectively. Indeed, $\PopL(a)$ pushes $a$ on $L$ if at least one element of $L$
 has been discarded and $a$ is not already on $L$, whereas $\PopR$ discards elements, but never pushes $c$ on $R$.
\er

Algorithm $\LR$-Search (see Algorithm \ref{algo:LR}) works intuitively as follows.
For each pair $(s,s+1$), the pair $[\bs,\Bs]$ of bounding values means that each
common interval $(a..c)$ containing $s$ and $s+1$ must satisfy the {\em bounding condition} 
$a\leq \bs < \Bs\leq  c$. The $\LR$-stack,  initially empty, stores on $L$ (respectively on $R$) the candidates for 
the left endpoint $a$ (respectively right endpoint $c$) of a common interval $(a..c)$, in such
a way that, after the execution of the {\bf for} loop (step 4) for a value $s$ with $a\leq s\leq c-1$, 
we have $c\in \Ra(a)$ iff $(a..c)$ satisfies all the bounding conditions previously imposed
with $s'$ such that $s\leq s'\leq c-1$. It is obvious (and understood) that if $(a..c)$ satisfies those conditions,
all intervals $(a'..c)$ with $a'<a$ do. When the execution of the {\bf for} loop considers $s=a$,
the bounding conditions are satisfied for all $s'$ with $a\leq s'\leq c-1$.

More precisely, we have the following theorem. Denote by $\Ra^s(a)$ the value of $\Ra(a)$ at the end of 
step 9 in the execution of the {\bf for} loop for $s$, for each $a$ on $L$.

\bthm  \cite{IR2013}  Assuming the \out$\,\,$procedure does not change the state of the $\LR$-stack, the 
set $Z$ defined as 

$$Z~:~=\bigcup_{1\leq s<p}\{(s..u)\, |\, u\in \Ra^s(s)\}$$
computed by $\LR$-Search is the set of all common intervals $(s..u)$ of $\PK$
 satisfying  

\begin{equation} 
s=\bs=\min\{l_w\,|\, s\leq w\leq u-1\}
\label{eq:bs}
\end{equation}

\begin{equation} 
u=r_{u-1}=\max\{r_w\,|\, s\leq w\leq u-1\}
\label{eq:Bu}
\end{equation}

\label{thm:B}

\ethm

\subsection{Setting the parameters}

With the aim of computing the common intervals of $\Id$ and $\pi$ that are valid with respect to $F$ and 
$\Phi$, we set the parameters $\BL$ and $\BR$ as follows. 

For each $s\in\{1, 2, \ldots, p-1\}$, let $v_s$ be as large as possible such that
$\Phi(v_s)$ is defined and

\begin{equation}
\pi[v_s..\Phi(v_s)]\, \mbox{contains both}\,   s,s+1
\label{eq:small}
\end{equation} 

Exactly one pair satisfies this condition, since $\Phi(1)=p$ and $\Phi$ is a (partial) function.  Now, let 

\begin{equation}
x_s=\max\{\pi^{-1}(s), \pi^{-1}(s+1)\}
\label{eq:xps}
\end{equation}

\noindent i.e. $x_s$ is the position of the rightmost element between $s$ and $s+1$ on $\pi$. Note 
that $x_s\leq \Phi(v_s)$, and let for all $s\in\{1, 2, \ldots, p-1\}$:

$$\bs=\min \pi[v_s..x_s] \hspace*{1cm} \Bs=\max \pi[v_s..x_s],$$

\noindent where $\min \pi[v_s..x_s]$ (respectively $\max \pi[v_s..x_s]$) denotes the minimum 
(respectively maximum) value in $\pi[v_s..x_s]$.

\begin{fait}
Assuming the \out$\,\,$procedure does not change the state of the $\LR$-stack, the set $Z$ defined as 

$$Z~:~=\cup_{1\leq s<p}\{(s..u)\, |\, u\in \Ra^s(s)\}$$
computed by $\LR$-Search with the settings $\BL,\BR$ above is the set of common intervals of
$\Id$ and $\pi$ that are valid with respect to $\Phi$.
\label{claim:A}
\end{fait}

{\bf Proof.} By Theorem \ref{thm:B}, we have to show that the set $C$ of common intervals satisfying
conditions (\ref{eq:bs}) and (\ref{eq:Bu}) is exactly the set $V$ of common intervals that are valid with 
respect to $\Phi$.

$''C\subseteq V''$: Let $(s..u)\in C$, and let $[y,z]$ be the location of $(s..u)$ in $\pi$.
Then by (\ref{eq:bs}) and (\ref{eq:Bu}):

\begin{equation}
s=\bs=\min\{l_w\,|\, s\leq w\leq u-1\}=\min\{ \min\pi[v_w..x_w]\,|\, s\leq w\leq u-1\},  
\label{eq:bsl}
\end{equation}
\begin{equation}
u=r_{u-1}=\max\{r_w\,|\, s\leq w\leq u-1\}=\max\{ \max\pi[v_w..x_w]\,|\, s\leq w\leq u-1\}.  
\label{eq:Bul}
\end{equation}

We first show that $y=\min\{v_w\, |\, s\leq w\leq u-1\}$. Assume a contrario that this is not the case. 
By equations (\ref{eq:bsl}) and (\ref{eq:Bul}) we deduce that $\pi[v_w]\in (s..u)$, for all $w$
with $s\leq w\leq u-1$. Thus 
$y<\min\{v_w\, |\, s\leq w\leq u-1\}$, since $y$ is the left endpoint of the location of $(s..u)$ on $\pi$. 
Now, with $\pi[y]\in (s..u)$, we deduce that:
\medskip

\begin{tabular}{ll}
$\bullet$& \begin{minipage}[t]{12cm} either $\pi[y]<u$ and thus $\pi[y]$ is one of the values $w$
with $s\leq w\leq u-1$. By equations (\ref{eq:small}) and (\ref{eq:xps}), $\pi[y]$ belongs
to $\pi[v_{\pi[y]}.. x_{\pi[y]}]$, therefore $y\geq v_{\pi[y]}$, a contradiction.\medskip\end{minipage}\\

$\bullet$& \begin{minipage}[t]{12cm} or $\pi[y]=u$ and thus $\pi[y]-1$ is one of the values $w$
with $s\leq w\leq u-1$. By equations (\ref{eq:small}) and (\ref{eq:xps}), $\pi[y]$ belongs
to $\pi[v_{\pi[y]-1}.. x_{\pi[y]-1}]$, therefore $y\geq v_{\pi[y]-1}$, a contradiction.\medskip\end{minipage}
\end{tabular}

Thus $y=\min\{v_w\, |\, s\leq w\leq u-1\}$, i.e. $y=v_{f}$ with $s\leq f\leq u-1$  and thus $\Phi(y)$ is defined.  
In a similar way, we are able to show that $z=\max\{x_w\, |\, s\leq w\leq u-1\}$, i.e. $z=x_{g}$ 
with $s\leq g\leq u-1$ and thus there exists $v_g$ such that $v_g\leq x_g\leq \Phi(v_g)$.

In the case where $\Phi(y)\geq z$, we have that $[y,z]$ is valid with respect to $\Phi$ and the proof is
finished. Let us show that one cannot have $\Phi(y)<z$. Indeed, if this is true, let $w$ with $s\leq w\leq u-1$
be such that one of $w,w+1$ is in $\pi[y..\Phi(y)]$ , and the
other one is in $\pi[\Phi(y)+1.. z]$. Such a value $w$ must exist, since $f,f+1$ are in $\pi[y..\Phi(y)]$, 
whereas at least one of $g,g+1$ is in $\pi[\Phi(y)+1.. z]$ by the definition of $x_g(=z)$.
Thus at least one integer $w$ with $min(f,g)\leq w\leq max(f,g)$ satisfies the required condition. But then $v_w\leq \Phi(y)$ and $x_w> \Phi(y)$, implying that $\Phi(v_w)\geq x_w>\Phi(y)$. But 
by Remark \ref{rmk:nested} the orders $\pi[y..\Phi(y)]$ and $\pi[v_w.. \Phi(v_w)]$ should dominate each other, and this is
impossible, since $v_w>v_f$ (the equality is forbidden by the definition of a function since
$\Phi(v_w)\neq \Phi(v_f)$), and $v_w<v_f$ means
that $v_w<y$ and this is impossible, since we proved $y=\min\{v_w\, |\, s\leq w\leq u-1\}$.

''$V\subseteq C$'': Let $(s..u)$ be a common interval of $\Id$ and $\pi$ which is valid with respect to $\Phi$.
Then the location $[y,z]$ of $(s..u)$ on $\pi$ satisfies $z\leq \Phi(y)$.
We have to show that equations (\ref{eq:bs}) and (\ref{eq:Bu}) hold.

For each $w$ with $s\leq w\leq u-1$, the interval $\pi[v_w..x_w]$ is the minimum interval which is valid
with respect to $\Phi$ that contains both $w$ and $w+1$. Since
$\pi[y..z]$ also contains $w,w+1$, we deduce from (\ref{eq:xps}) that $y\leq v_w\leq x_w\leq z$,
implying that

\begin{equation}
\min \pi[y..z]\leq \min \pi[v_w..x_w]\,  \mbox{and}\, \max \pi[y..z]\geq \max \pi[v_w..x_w],
\label{eq:bB}
\end{equation}

\noindent Now, over all $w$ we deduce:

\begin{equation}
\min \pi[y..z]\leq \min\{\min \pi[v_w..x_w]\, |\, s\leq w\leq u-1\}=\min\{ l_w\, |\, s\leq w\leq u-1\}
\label{eq:bsn}
\end{equation}
\begin{equation}
\max \pi[y..z]\geq \max\{\max \pi[v_w..x_w]\, |\, s\leq w\leq u-1\}=\max\{ r_w\, |\, s\leq w\leq u-1\}
\label{eq:Bun}
\end{equation}

Furthermore, we have $\min \pi[y..z]=s$ and $\max \pi[y..z]=u$ since $[y,z]$ is the location of $(u..s)$ on
$\pi$. Moreover, we have $s=l_s$, otherwise by the definition of $l_s$ we only have the possibility 
$l_s<s$, contradicting equation (\ref{eq:bsn}).
Similarly, $u=r_u$ and the proof is finished. $\Box$
\bigskip

Consider now the filtering procedure in Algorithm \ref{algo:filterR}, which chooses among all common intervals 
produced by $\LR$-Search, and which
are valid w.r.t. $\Phi$, those that are also valid w.r.t. $F$.

\begin{algorithm}[t,boxed]
\caption{The \out$\,\,$algorithm for selecting common intervals that are valid w.r.t. $F$ and $\Phi$} 
\begin{algorithmic}[1]
\REQUIRE Pointers $\Rt(s),\Rb(s)$ to the first and last element of $\Ra(s)$ (possibly equal to $nil$)
\ENSURE  All common intervals $(s..u)$ of $\Id$ and $\pi$, with fixed $s$, that are valid w.r.t. $F$ and $\Phi$

\IF{$F(s)$ is defined and $\Rt(s)\neq nil$}
\STATE $u^{\top}\leftarrow $ the target of $\Rt(s)$; $u^{\bot}\leftarrow $ the target of $\Rb(s)$
\STATE $u\leftarrow u^{\top}$
\WHILE{$u\leq u^{\bot}$ and $u\leq F(s)$}
\STATE Output the interval $(s..u)$
\STATE $u\leftarrow next(u)$ \hfill {\sl //or $p+1$ if $\next(u)$ does not exist}
\ENDWHILE
\ENDIF
\end{algorithmic}
\label{algo:filterR}
\end{algorithm}

\bthm
Algorithm $\LR$-Search with settings $l_s, r_s$ and \out\, in Algorithm \ref{algo:filterR}  outputs the common intervals of
$\Id$ and $\pi$ which are valid w.r.t. $F$ and $\Phi$. The algorithm runs in  $O(p+N_{\Id,\pi})$ time,
where  $N_{\Id,\pi}$ is the number of such common intervals, and uses $O(p)$ space.
\label{thm:common}
\ethm

{\bf Proof.} Claim \ref{claim:A} guarantees that the intervals $(s..u)$ with $u\in \Ra^s(s)$ are
exactly the common intervals of $\Id$ and $\pi$ which are valid with respect to $\Phi$. Recall
that $\Ra^s(s)$ is $\Ra(s)$ at the end of step 9 during the execution of the {\bf for} loop for $s$ in the
$\LR$-Search algorithm,
that is, exactly the set $\Ra(s)$ considered by \out. Moreover, it is easy to see that the
\out\, procedure selects between these intervals $(s..u)$ those for which $F(s)$ is defined (step 1)
and $u\leq F(s)$ (step 4). Only these intervals are output, so that the correctness
of the algorithm is proved.

The running time and memory space requirements are both in $O(p)$, when \out\, is left apart and 
$l_s, r_s$ are supposed already computed, as proved in \cite{IR2013} (case of $K=2$ permutations).  
This is done by implementing the two stacks $L$ and $R$ of the $\LR$-stack as lists, so as to insure that $\PopL, \PopR$
are performed in linear time with respect to the number of elements removed from $L$ and $R$ respectively.
Furthermore, the running time of \out\, is proportional with the number of output intervals, 
and there are no supplementary space requirements. Note that $\Rt,\Rb$ are easily computed when
$\PopL, \PopR$ and $\PushLR$ are performed.

It remains to show that $l_s$ and $r_s$ may be computed in global $O(p)$ time and space, for all values of $s$.
To this end,  algorithm ComputeV($\pi$) in Algorithm \ref{algo:v} efficiently computes the values $v_s$ as shown below.
The values $x_s$ are easy to compute.
Once this is done, computing $\bs$ (and similarly $\Bs$) for each $s$ is solving a problem known as the
{\sc Range Minimum Query} problem \cite{berkman1993recursive, bender2000lca}
for the set of pairs $Q=\{(v_s,x_s)\, |\, s=1, 2, \ldots, p-1\}$. This takes $O(1)$ for each pair once a $O(p)$  preprocessing of $\pi$ is done \cite{berkman1993recursive, bender2000lca}. Alternatively, one may use the simplest
algorithm in \cite{IR2013}, based on $\LR$-stacks and which also guarantees a $O(p)$ global time.

In algorithm ComputeV($\pi$), every pair $(w,\Phi(w))$ is recorded on $\rho$ as a pair of separators (steps 2-4), 
defining slices similarly to Section \ref{sect:findpos} (see Exemple \ref{ex:computeV}). For each pair of
consecutive separators in $\rho$, the positions in $\pi$ of the elements located between these separators
form a slice. Its identification number is the smallest value it contains. Slices are sets of consecutive values, 
implemented  again as a Union-Find structure limited to unions between neighboring sets,
using the structure proposed in \cite{itai2006linear} which allows to perform $u$ unions and $f$ finds (here
called $FindSet$)
in $O(u+f)$ time. Then, the elements of $\rho$ (of length in $O(p)$) are successively considered
and produce either two set unions (step 8) or at most two calls of $FindSet$ (steps 11-12). Consequently, the
running time of the algorithm is linear with respect to the size of $\pi$, and is thus in $O(p)$.

An invariant of the algorithm, easy to prove, is that for each element $s$ of $\pi$ (which thus belongs to $\rho$), 
the closest left $w$-separator still present on its left in $\rho$ is the best current candidate for $v_s$ and $v_{s-1}$,
whenever $s+1$ and/or $s-1$ have not been seen yet. Each right $w$-separator situated between $s$ and
one of these values, found during the search in step 6, produces the removal of both $w$-separators and thus
the update of the candidate. When $s+1$ or $s-1$ is found (steps 11-12), the best current
candidate is the sought value. The correctness of the algorithm follows. $\Box$

\bex
With $\pi=5\, 3\, 1\, 4\, 2\, 6$ and $\Phi$ defined as $\Phi(1)=6$ and $\Phi(3)=5$, the {\bf for} loop in
steps 2-4 of algorithm  ComputeV($\pi$) yields $\rho=[_1\, 5\, 3\, [_3\, 1\, 4\, 2\, ]_3\, 6\, ]_1$,
with slices $\{1,2\}$ (the positions of 5 and 3 in $\pi$), $\{3, 4, 5\}$ (the positions of $1, 4$ and 2 in $\pi$), 
and $\{6\}$ (the position of $6$ in $\pi$). Each slice is a set identified by its smallest element. The
{\bf for} loop in steps 6-14 starts with $j=1$. As $\rho[1]$ is a left separator, nothing happens. For $j=2$, the
condition in step 10 is true, as $\rho[2]=5$. However, $\rho^{-1}(4)=6>2$ in step 11 and $\rho^{-1}(6)=9>2$ in step 12,
thus no value is set in these steps. For $j=3,4$ and $5$, similar situations occur. Now, for $j=6$ we have $\rho[j]=4$,
and the conditions in steps 11 and 12 are both true since $\rho^{-1}(3)=3<6$ and $\rho^{-1}(5)=2<6$. Then
$v_3\leftarrow FindSet(\pi^{-1}(3))$, which is $FindSet(2)$ i.e. 1, since the slice $\{1,2\}$ has identification number 1. Similarly, $v_4\leftarrow FindSet(\pi^{-1}(5))$ which is also 1. Continuing
with $j=7$, we have $\rho[j]=2$ which gives $v_1\leftarrow FindSet(\pi^{-1}(1))$ (which is 3) and $v_2\leftarrow 
FindSet(\pi^{-1}(3))$ (which is 1). When $j=8$, the condition in step 7 is true with $w=3$ so  that the left and right 
$3$-separators are removed, yielding $\rho=[_1\, 5\, 3\, 1\, 4\, 2\, 6\, ]_1$ with the unique slice $\{1,2, 3, 4, 5, 6\}$
(the positions of $5, 3, 1, 4, 2, 6$). With $j=9$ we obtain $v_5\leftarrow 1$.  
\label{ex:computeV}
 \eex

Algorithm \ref{algo:main} is now complete.

\begin{algorithm}[t,boxed]
\caption{ComputeV($\pi$)}
\begin{algorithmic}[1]
\REQUIRE An order $\pi$ and its dominating function $\Phi$
\ENSURE  For each $s$ with $1\leq s<p$, the value $v_s$.

\STATE $\rho\leftarrow \pi$
\FOR{each $w$ such that $\Phi(w)$ is defined}
\STATE insert a left $w$-separator $[_w$ immediately before the value $\pi[w]$ and a right $w$-separator $]_w$ immediately after the value $\pi[\Phi(w)]$ in $\rho$
\ENDFOR
\STATE let each slice in $\rho$ be a set of positions in $\pi$, identified by its smallest element
\FOR{$j=1$ to $||\rho||$}
\IF{$\rho[j]$ is a right separator $]_w$}
\STATE remove the separators $[_w$ and $]_w$ from $\rho$\hfill{\sl //virtually; in the facts, update slices and sets}
\ELSE
\IF{$\rho[j]$ is a value from $\pi$}
\STATE {\bf if} $\rho[j]-1\geq 1$ and $\rho^{-1}(\rho[j]-1)<j$ {\bf then} $v_{\rho[j]-1}\leftarrow FindSet(\pi^{-1}(\rho[j]-1))$ {\bf end if}
\STATE {\bf if} $\rho[j]+1\leq p$ and $\rho^{-1}(\rho[j]+1)<j$ {\bf then} $v_{\rho[j]}\leftarrow FindSet(\pi^{-1}(\rho[j]+1))$ {\bf end if}
\ENDIF
\ENDIF
\ENDFOR
\end{algorithmic}
\label{algo:v}
\end{algorithm}

\section{Conclusion} \label{sect:conclusion}

In this paper, we proposed an alternative view of the common interval searching in sequences, obtained by
reducing both sequences to orders (which are permutations), and factorizing the search for common intervals
by grouping several orders into a unique order, called a dominating order. In order to insure low time
and space requirements, we had to avoid storing the orders, but also rebuilding them with the algorithm
from scratch in Section \ref{sect:find}, which was too time consuming to be used in steps 3 and 4 of the
main algorithm. Then, we used minimal informations computed by the algorithm from scratch in order to propose
in Section \ref{sect:retrieve} a much more efficient algorithm to generate the orders. Finally, we
used the parameterizable algorithm in \cite{IR2013} to give an algorithm able to find common intervals
with constrained endpoints, which completes our search for common intervals in sequences.

To solve the $(T,S)$-{\sc Common Interval Searching} problem, we reduced it to a problem called
$(P,S)$-{\sc Guided Common Intervals Searching}, where $P$ is a permutation on $p$ elements. We proposed a
$O(q_2n_2+N_{P,S})$-time algorithm for solving this problem. However, an improved running time of  
$O(n_2+N_{P,S})$ for solving this problem would lead to a $O(q_1n_1+q_2n_2+N)$ algorithm for the case of two sequences,
improving the existing $O(n_1n_2)$ algorithms.

\bibliographystyle{plain}
\bibliography{biblio}
\end{document}